\documentclass[preprintnumbers,superscriptaddress,nofootinbib,aps,prd,floatfix,twocolumn]{revtex4-2}
\pdfoutput=1
\usepackage{enumerate}
\usepackage{amsmath,amssymb}
\usepackage{graphicx}
\usepackage{slashed}
\usepackage{xspace,slashed}
\usepackage{float}
\usepackage[dvipsnames]{xcolor}
\usepackage{hyperref}
\hypersetup{colorlinks=true, citecolor=Blue, urlcolor=Blue, linkcolor=Blue}
\usepackage[normalem]{ulem}
\usepackage{subfigure,orcidlink}
\usepackage{multirow,array}
\usepackage{braket}
\usepackage{marginnote}
\usepackage{eurosym}
\usepackage{tabularray}
\UseTblrLibrary{booktabs}
\usepackage{bbding}
\begin{document}
\title{Does thermal leptogenesis in a canonical seesaw rely on initial memory?}

\author{Partha Kumar Paul$^{\orcidlink{https://orcid.org/0000-0002-9107-5635}}$}
\email{ph22resch11012@iith.ac.in}
\affiliation{Department of Physics, Indian Institute of Technology Hyderabad, Kandi, Sangareddy, Telangana-502285, India.}

\author{Narendra Sahu$^{\orcidlink{https://orcid.org/0000-0002-9675-0484}}$}
\email{nsahu@phy.iith.ac.in}
\affiliation{Department of Physics, Indian Institute of Technology Hyderabad, Kandi, Sangareddy, Telangana-502285, India.}
    
\author{Shashwat Sharma$^{\orcidlink{https://orcid.org/0009-0002-2266-0467}}$}
\email{ph23resch11016@iith.ac.in}
\affiliation{Department of Physics, Indian Institute of Technology Hyderabad, Kandi, Sangareddy, Telangana-502285, India.}
  
\begin{abstract}
It is a common lore that in thermal leptogenesis within the type-I seesaw framework and a hierarchical spectrum of heavy right-handed neutrinos (RHNs), the CP-violating, out-of-equilibrium decay of the lightest RHN ($N_1$) is the only relevant source of the final $B-L$ asymmetry, since any asymmetry produced by the heavier RHNs is expected to be erased by subsequent $N_1$-mediated washout processes.
In this work, we revisit this assumption by solving the density-matrix equations, including decay, inverse decay, and relevant scattering processes, and by fully accounting for flavor-projection effects induced by the Yukawa coupling structure. We show that the asymmetries generated by the heavier RHNs ($N_2$ and $N_3$) generally possess components that are misaligned in flavor space with respect to $N_1$, resulting in a partially protected contribution that survives the $N_1$ washout.
Unlike the conventional picture of $N_2$-dominated leptogenesis, this memory effect arises even when $N_1$ remains dynamically relevant and cannot be captured within the classical Boltzmann framework. Furthermore, imposing consistency with low-energy neutrino mass and mixing data, we find that at most one RHN can lie in the weak washout regime, which naturally divides the parameter space into four distinct dynamical regimes. We systematically quantify the memory effect in each regime and demonstrate that it can significantly modify the final $B-L$ asymmetry. We find that including projection effects can indeed extend the viable parameter space into the sensitivity range of neutrinoless double beta decay experiments.
\end{abstract}

\date{\today}
\maketitle
\noindent
\section{Introduction}
\label{sec:intro}

One of the most profound mysteries in cosmology is the observed baryon asymmetry of the Universe, which is the imbalance between matter and antimatter. The early Universe should have created equal amounts of matter and antimatter, but observations show that matter dominated over antimatter. This asymmetry is quantified as the baryon-to-photon ratio: 
$\eta_B = \frac{n_B - n_{\bar{B}}}{n_{\gamma}} = (6.14 \pm 0.04) \times 10^{-10}$
as determined by the cosmic microwave background \cite{Planck:2018vyg} and the big bang nucleosynthesis \cite{Fields:2004cb,ParticleDataGroup:2024cfk, Cyburt:2004yc, Steigman:2005uz}. The standard model (SM) of particle physics cannot account for this observed imbalance between matter and antimatter. 

Thermal leptogenesis in a canonical seesaw provides a simple yet compelling mechanism for explaining the observed baryon asymmetry of the Universe \cite{Fukugita:1986hr,PhysRevD.45.455,PhysRevD.46.5331,Flanz:1994yx,Davidson:2008bu,Buchmuller:2004nz, Barbieri:1999ma,Pilaftsis:2003gt,Pilaftsis:2005rv,Nardi:2006fx}. It connects the matter-antimatter asymmetry with the origin of non-zero neutrino mass as revealed by oscillation experiments \cite{deSalas:2017kay,T2K:2015sqm,IceCube:2015fuw,Bahcall:2004mz, Super-Kamiokande:2001bfk,KamLAND:2002uet} via the type-I seesaw framework \cite{Minkowski:1977sc,Gell-Mann:1979vob,Sawada:1979dis,PhysRevLett.44.912,Schechter:1980gr, Valle:2015pba}. This mechanism is based on the $CP$ violating out-of-equilibrium decay of heavy right-handed neutrinos (RHNs) to SM leptons and Higgs, producing a net lepton asymmetry which finally gets converted to the baryon asymmetry via the electroweak (EW) sphalerons \cite{Sakharov:1967dj,Kuzmin:1985mm}. A common assumption in this framework is that the lightest RHN, $N_1$ is the primary contributor to the final $B-L$ asymmetry, as any asymmetries generated by the decays of heavier RHNs ($N_2,N_3$) are presumed to be erased by the lepton number violating processes mediated by $N_1$ \cite{Buchmuller:2004nz}. This assumption simplifies the analysis of thermal leptogenesis, effectively decoupling the dynamics of $N_1$ from $N_2$ and $N_3$. However, as we delineate in this paper, the validity of this assumption is not always correct.

In this paper, we revisit the validity of this assumption by considering four different cases. We saw that, to satisfy the neutrino masses and mixing given by the low-energy neutrino oscillation data, at most one of the RHNs can be in the weak washout regime. Therefore, the four different cases can be: I) All the RHNs are in the strong washout regime, II) $N_1$ is in the weak washout regime while $N_2$ and $N_3$ are in the strong washout regime, III) $N_2$ is in the weak washout regime while $N_1$ and $N_3$ are in the strong washout regime, and finally IV) $N_3$ is in the weak washout regime while $N_1$ and $N_2$ are in the strong washout regime. In all the above four cases, we use the density matrix formalism to demonstrate the percentage survival of the pre-existing asymmetries produced by the heavier RHNs.

We assume that thermal leptogenesis occurs at temperatures $T\gtrsim10^{12} \rm{GeV}$, where the charged-lepton Yukawa interactions are out of equilibrium. In this regime, the lepton doublets produced in the decays of RHNs are usually assumed to be indistinguishable, and hence all generated asymmetries are considered to be aligned in flavor space\cite{Plumacher:1996kc}. This leads to a complete erasure of the pre-existing asymmetries produced by $N_2$ and $N_3$ via lepton number-violating processes mediated by $N_1$.
However, this assumption of perfect alignment among the asymmetries produced by different RHNs is not generally true and strongly depends on the Yukawa coupling texture. In reality, the lepton states produced by the decays of distinct RHNs need not be identical—each RHN couples to a different linear combination of lepton flavors ($\ket{l_i}=\sum_\alpha\mathcal{C}_{i\alpha}\ket{l_\alpha}$ where $\alpha=e,\mu,\tau $) \cite{Barbieri:1999ma,Vives:2005ra,Engelhard:2006yg,BariDensityMatrix,N2leptogenesisprojection,Leptogenesisin2RHNModel,JessicaTurner,BariReview,DiBari:2005st,Hahn-Woernle:2009okg, ReFiorentin:2015rrc}. Consequently, the asymmetries generated by $N_2$ and $N_3$ can have components that are orthogonal to the direction of $N_1$-decay and are not erased by the lepton number violating interactions of $N_1$, while those parallel to the direction of $N_1$-decay can be erased.
This leads to a protected portion of the asymmetries generated by $N_{2,3}$ that survive and contribute to the final $B-L$ asymmetry. We refer to the percentage survival of the pre-existing asymmetry as ``\textit{memory effect}". Such effects cannot be captured within the classical Boltzmann approach, which tracks only the total lepton number density. To consistently describe the evolution of coherence among different heavy-neutrino flavor states and the corresponding asymmetries, we use the density matrix formalism. We solve the density matrix equation for the four washout regimes discussed in the previous paragraph and delineate the memory effect in all cases.

The paper is organized as follows. In section \ref{sec:leptointype1}, we briefly recapitulate the type-I seesaw mechanism, while in section \ref{sec:leptogenesis}, we show the details of the density matrix formalism for leptogenesis. We discuss the results in section \ref{sec:result} by comparing the memory effect ($\delta$) with respect to the mass ratios ($M_2/M_1$ and $M_3/M_1$), considering decays, inverse decays, and scatterings. We compare the numerical solution of the full density matrix equations with the analytical approximation in section \ref{sec:x}. In section  \ref{sec:nu0bb}, we discuss the importance of projection effects in neutrinoless double beta decay. We finally conclude in section \ref{sec:conclu} by summarizing our results and discussing the impact of heavier RHNs ($N_2$ and $N_3$) on the final $B-L$ asymmetry.

\section{type-I Seesaw}\label{sec:leptointype1}

In the canonical type-I seesaw, the SM is extended with three RHNs, $N_1,N_2,N_3$, which are singlets under the SM gauge group ($SU(3)_C \times SU(2)_L \times U(1)_Y$). The Lagrangian responsible for generating lepton asymmetry as well as neutrino mass is given as \footnote{We have suppressed the generation indices.}
\begin{equation}\label{eq:type1}
   - \mathcal{L}^{{\rm type-I}} \supset \frac{1}{2}M_R \bar{N^c}N + Y \bar{l} \tilde{H} N + {\rm H.c.}
\end{equation}
Here, $\tilde{H}=i\sigma_2 H^*$ where $H$ is the Higgs doublet and $l=\begin{pmatrix}
    \nu_L \\ l_L
\end{pmatrix}$ is the lepton doublet.
From Eq.(\ref{eq:type1}) we can write the $6\times 6$ neutral fermion mass matrix as
\begin{equation}
    \begin{pmatrix}
        \bar{\nu}_L^c & \bar{N}_R
    \end{pmatrix}
    \begin{pmatrix}
        m_L & m_D \\ m_D^T & M_R
    \end{pmatrix}
    \begin{pmatrix}
        \nu_L \\ N_R^c
    \end{pmatrix}
    +{\rm H.c.}
\end{equation}
where $m_D = \frac{Yv}{\sqrt{2}}$ ($v$ is the Higgs vacuum expectation value (vev)). Thus, for $M_R\gg vY/\sqrt{2}$, we can write
\begin{equation}
    m_\nu= - m_D M_R^{-1} m_D^T =-\frac{v^2}{2} Y M_R^{-1} Y^T.\label{eq:mnu}
\end{equation}
Thus, the large RHN masses give an origin to the tiny masses of the active neutrinos via the type-I seesaw mechanism. The light neutrino mass matrix ($m_\nu$) can be diagonalized using the unitary Pontecorvo–Maki–Nakagawa–Sakata (PMNS) matrix $U$
\begin{equation}
    D_m = U^T m_\nu U,\label{eq:Dm0}
\end{equation}
where $D_m = {\rm diag}(m_1,m_2,m_3)$. Using Eq.(\ref{eq:mnu}) in Eq.(\ref{eq:Dm0}), we get
\begin{equation}\label{eq:Dm}
    D_m=-\frac{v^2}{2}U^T Y M_R^{-1} Y^T U.
\end{equation}
Without loss of generality, we assume the heavy RHNs to be in a diagonal mass basis. We then denote $M_R = D_M = {\rm diag}(M_1, M_2, M_3)$, thus from Eq.(\ref{eq:Dm}) we can write
\begin{equation}
     -\frac{v^2}{2} (D_{\sqrt{m^-1}} U^T Y D_{\sqrt{M^{-1}}}) (D_{\sqrt{M^{-1}}} Y^T U D_{\sqrt{m^{-1}}}) = I_{3\times 3}.
\end{equation}
Here, the notation $D_{\sqrt{A}}$ means $\sqrt{D_{A}}$. Since, $(D_{\sqrt{M^{-1}}} Y^T U D_{\sqrt{m^{-1}}})^T = (D_{\sqrt{m^{-1}}} U^T Y D_{\sqrt{M^{-1}}})$, we can write
\begin{equation}
    \frac{-iv}{\sqrt{2}}(D_{\sqrt{m^{-1}}} U^T Y D_{\sqrt{M^{-1}}}) = R^T,\label{eq:RT}
\end{equation}
where, $R$ is a complex orthogonal matrix with $R^T R = I$. Thus, using Eq.(\ref{eq:RT}), we can write the Yukawa coupling matrix as
\begin{equation}
    Y = \frac{i \sqrt{2}}{v} U^* D_{\sqrt{m}} R^T D_{\sqrt{M}}.\label{eq:casasiba}
\end{equation}
This is known as the Casas-Ibarra Parameterization of the Yukawa coupling matrix \cite{Casas:2001sr}.
This implies that larger values of RHNs' masses lead to larger Yukawa couplings, holding all other parameters fixed. The PMNS matrix $U$ is given by \cite{ParticleDataGroup:2024cfk}
{{\begin{eqnarray} 
U &=& 
    \begin{pmatrix}
        c_{12} c_{13} & s_{12} c_{13} & s_{13} e^{-\iota \delta} \\ -s_{12} c_{23} - c_{12} s_{13} s_{23} e^{\iota \delta} & c_{12} c_{23} - s_{12} s_{13} s_{23} e^{\iota \delta} & c_{13} s_{23} \\ s_{12}s_{23} - c_{12} s_{13} c_{23} e^{\iota \delta} & -c_{12} s_{23} - s_{12} s_{13} c_{23} e^{\iota \delta} & c_{13} c_{23}
    \end{pmatrix}\nonumber\\&&.
    \begin{pmatrix}
        e^{\iota \eta_1} & 0 & 0 \\ 0 & e^{\iota \eta_2} & 0 \\ 0 & 0 & 1
    \end{pmatrix},
\end{eqnarray}}}
where $c_{ij} = \cos \theta_{ij}$ and $s_{ij} = \sin \theta_{ij}$ and $\delta$ is the Dirac $CP$ phase and $\eta_{1,2}$ are the Majorana $CP$ phases. We randomly choose Majorana phases in the range ($0,2\pi$) and take the neutrino oscillation parameters to be in the $3\sigma$ range \cite{deSalas:2020pgw}. We also consider the mass of the lightest SM neutrino ($m_1$) to be zero. In the general case, the complex orthogonal matrix $R$ in Eq.(\ref{eq:casasiba}) can be described by 3 different complex angles ($\theta_i=x_i+iy
_i; i=1,2,3$) with a form of
\begin{eqnarray}
        R &=&
    \begin{pmatrix}
        \cos \theta_1 & -\sin \theta_1 & 0 \\
        \sin \theta_1 & \cos \theta_1 & 0\\
        0 & 0 & 1
    \end{pmatrix}
    \begin{pmatrix}
        \cos \theta_2 & 0 & \sin \theta_2 \\
        0 & 1 & 0\\
        -\sin \theta_2  & 0  & \cos \theta_2 
    \end{pmatrix}\nonumber\\&&
    \begin{pmatrix}
        1 & 0 & 0\\
        0 & \cos \theta_3  & -\sin \theta_3 \\
        0 & \sin \theta_3  & \cos \theta_3
        \end{pmatrix}\label{eq:Rmatrix}
\end{eqnarray}

Thus given, $\theta_1, \theta_2, \theta_3, M_{1},M_{2}, M_{3}$, one can derive a Yukawa coupling matrix using Eq.(\ref{eq:casasiba}).

\section{Density matrix formalism for sequential leptogenesis in type-I seesaw}\label{sec:leptogenesis}

\begin{figure}[H]
  \centering  \includegraphics[scale=0.8]{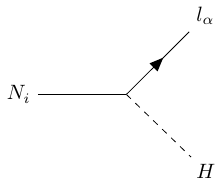}
  \includegraphics[scale=0.8]{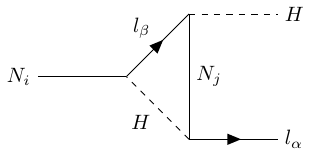}
  \includegraphics[scale=0.8]{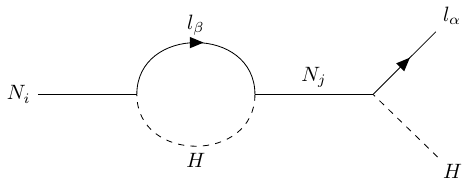}
  \caption{Tree and one-loop level decays of RHNs giving rise to nonzero $CP$ asymmetry.}  \label{fig:Cpasymmdiag}
\end{figure}

In the type-I seesaw, the Majorana nature of the heavy RHNs naturally leads to lepton number violation. In an expanding universe, the \textit{CP}-violating, out-of-equilibrium decays of the heavy RHNs can give rise to the lepton asymmetry \cite{Fukugita:1986hr,PhysRevD.45.455,PhysRevD.46.5331,Flanz:1994yx,Davidson:2008bu,Buchmuller:2004nz, Barbieri:1999ma}. The interference between the tree-level and one-loop decay diagrams (self-energy correction and vertex term) gives the required $CP$ asymmetry, as shown in Fig. \ref{fig:Cpasymmdiag}.

The $CP$ asymmetry parameter is defined as
\begin{equation}    \varepsilon_i=\frac{\Gamma(N_i \rightarrow l H)-\Gamma(N_i \rightarrow \bar{l} \bar{H})}{\Gamma(N_i \rightarrow l H)+\Gamma(N_i \rightarrow \bar{l} \bar{H})}
\end{equation}
Considering the interference of tree-level diagrams with one-loop diagrams, we have \cite{Davidson:2002qv,Davidson:2002qv}
\begin{equation}
    \varepsilon_i = -\frac{1}{8\pi}\sum_{j\neq i} \frac{Im[(Y^\dagger Y)_{ij}^2]}{(Y^\dagger Y)_{ii}}\bigg[f_v\bigg(\frac{M_j^2}{M_i^2}\bigg)+f_s\bigg(\frac{M_j^2}{M_i^2}\bigg)\bigg]
\end{equation}
where $f_v(x)$ comes from the one-loop vertex term and $f_s(x)$ comes from the one-loop self-energy term, and they are given by
\begin{equation}
\begin{split}
    &f_s(x)=\frac{\sqrt{x}}{1-x};\\
    &f_v(x)= \sqrt{x}\bigg[1-(1+x)\ln\bigg(\frac{1+x}{x}\bigg)\bigg].
\end{split}
\end{equation}

In standard (semi-classical) Boltzmann equations, we track the number densities of heavy neutrinos $N_i$ and the lepton doublets $l_\alpha$ ($\alpha=e,\mu,\tau$). 
At very high temperatures ($T\gtrsim10^{12} \rm{GeV}$), the charged-lepton Yukawa interactions are out-of-equilibrium, so all lepton flavors behave coherently, which is the flavor blind regime. If the heavy Majorana neutrino $N_i$ decays into a lepton/anti-lepton doublet at a temperature $T\gtrsim10^{12} \rm{GeV}$, the produced lepton/anti-lepton doublet $\ket{l_i}$ is a coherent superposition of the lepton flavor states $\ket{l_\alpha}$ given by \footnote{Without loss of generality we take $\ket{l_{\alpha}}(\alpha=e,\mu,\tau)$ to be an orthogonal basis.}
\begin{eqnarray}
    \ket{l_i}&=&\sum_\alpha c_{i\alpha} \ket{l_\alpha}, \quad c_{i\alpha}=\braket{l_\alpha|l_i},\nonumber\\
    \ket{\bar{l}_i}&=&\sum_\alpha \bar{c}_{i\alpha} \ket{\bar{l}_\alpha}, \quad \bar{c}_{i\alpha}=\braket{\bar{l}_\alpha|\bar{l}_i}
\end{eqnarray}
where the coefficients $c_{i\alpha}$ are determined by the Yukawa couplings $Y_{i\alpha}$ as $c_{i\alpha} =\frac{Y_{\alpha i}}{\sqrt{(Y^\dagger Y)_{ii}}}$, which satisfy $\sum_\alpha |c_{i\alpha}|^2=1$. 

\begin{figure}[H]
    \centering
    \includegraphics[scale=0.4]{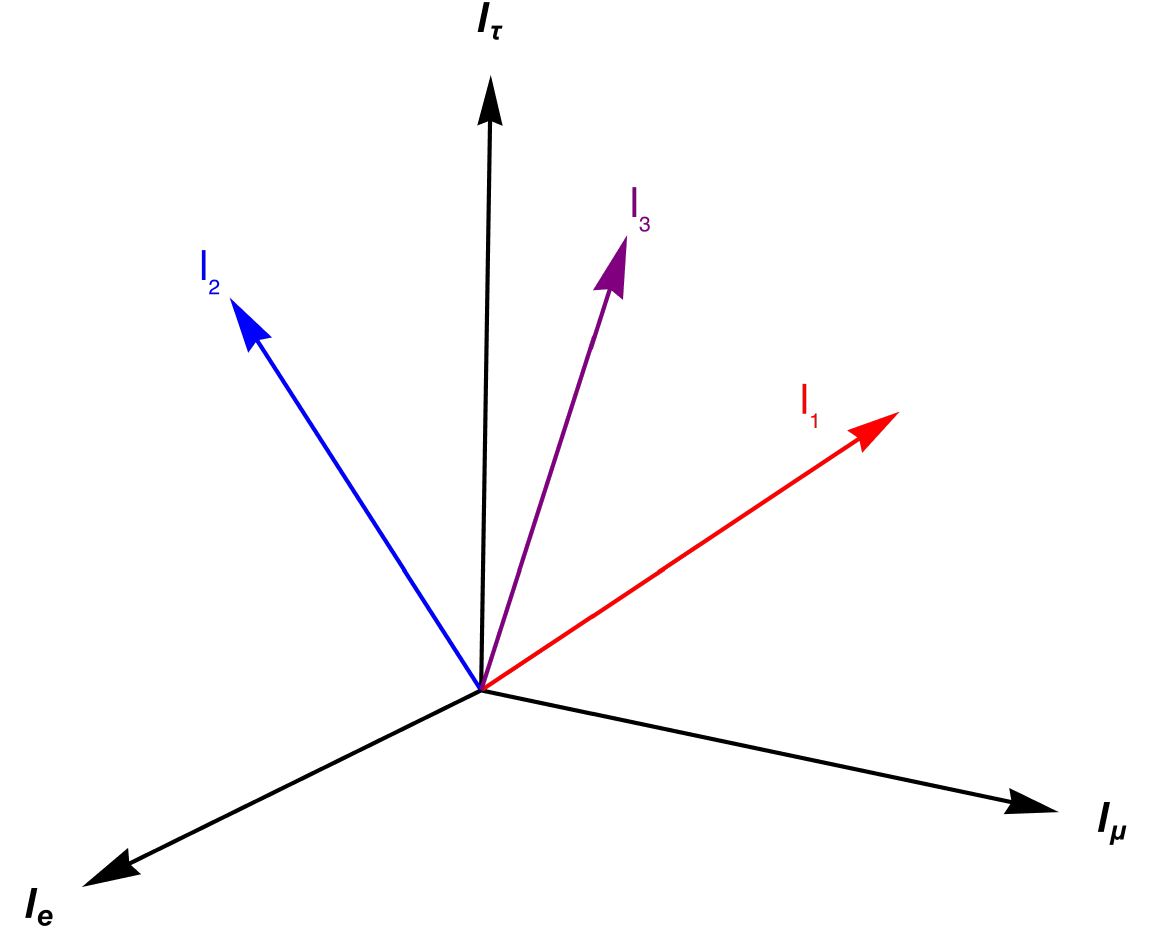}
    \caption{Figure illustrating the produced lepton $l_i$ from $N_i$ decays as ``flavor vector" in the flavor space.}
    \label{fig:3Dflavorbasis}
\end{figure}
\begin{figure}[H]
    \centering
    \includegraphics[scale=0.4]{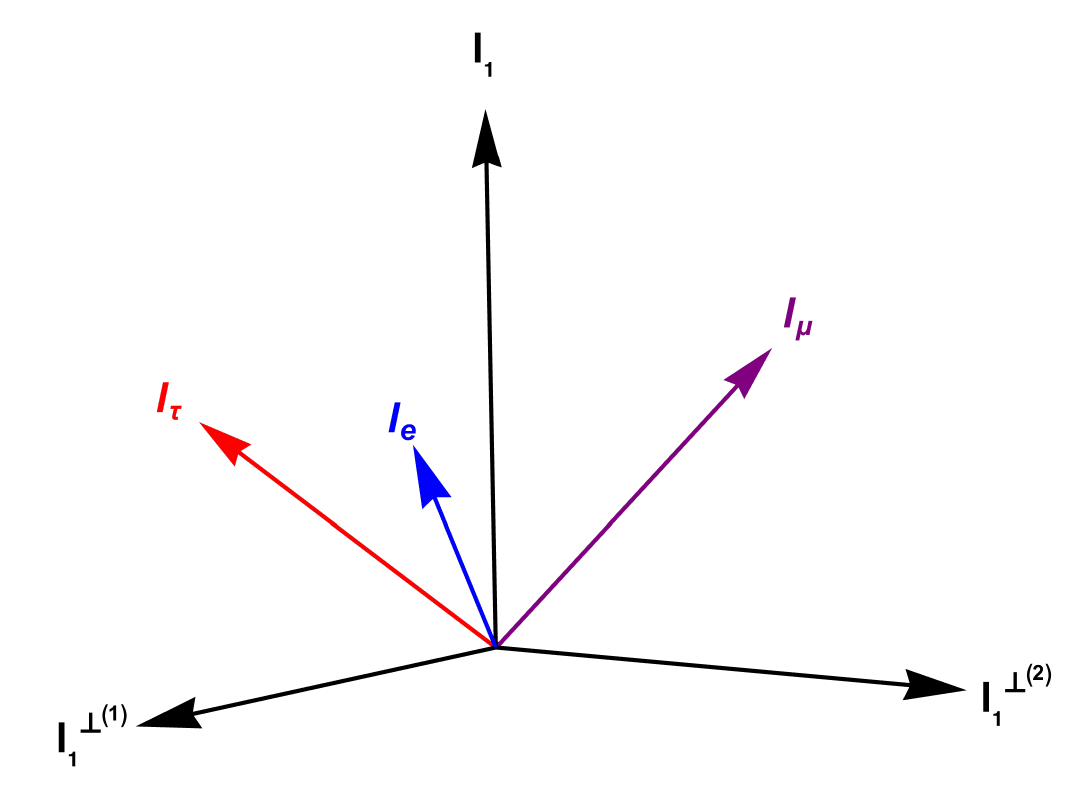}
    \caption{Figure illustrating the new basis $(l_i,l_i^{\perp^{(1)}},l_i^{\perp^{(2)}})$ defined in the flavor space.}
    \label{fig:3Dleptonbasis}
\end{figure}

Given that these ``vectors" have different coefficients, they are in general not parallel to each other in the flavor space, which is illustrated in Fig.\ref{fig:3Dflavorbasis}. Since they have some arbitrary alignment defined by the Yukawa coupling texture, it is allowed that $l_i$ has some component which is orthogonal to $l_j$ and is thus protected from inverse decay washout effects of $l_j$. This is called the projection effect.
In this paper, we consider the asymmetry produced by all three RHNs by taking into account the projection effects and compare our result with the asymmetry produced by $N_1$ decay only. Since $l_i$ are vectors in the flavor space, we can define 3 different orthogonal bases: $(l_i,l_i^{\perp^{(1)}},l_i^{\perp^{(2)}})$  where $i=1,2,3$ as shown in Fig.\ref{fig:3Dleptonbasis}. We conveniently refer to them as ``RHN flavor basis'' because $N_i$ are allowed to decay along the $l_i$ directions and are expressed as
\begin{eqnarray}
    \label{eq:leptonbasis}&&\ket{l_i}=c_{i\tau}\ket{l_\tau}+c_{i\mu}\ket{l_\mu}+c_{ie}\ket{l_e}\nonumber\\
    &&\ket{l_i^{\perp^{(1)}}}= \frac{0\ket{l_\tau}+c_{ie}^*\ket{l_\mu}-c_{i\mu}^*\ket{l_e}}{\sqrt{|c_{ie}|^2+|c_{i\mu}|^2}}\\
    &&\ket{l_i^{\perp^{(2)}}}=\frac{(1-|c_{i\tau}|^2)\ket{l_\tau}-c_{i\mu} c_{i\tau}^*\ket{l_\mu}-c_{ie} c_{i\tau}^*\ket{l_e}}{\sqrt{|c_{ie}|^2+|c_{i\mu}|^2}} \nonumber
\end{eqnarray}

Here, any asymmetry produced in the direction $l_i^{\perp^{1,2}}$ is protected from the washout effects in the $l_i$ direction (say for instance $l_i H \rightarrow N_i$).

Now, for the production of $l_i$ from $N_i \rightarrow l_iH$ interaction, we can define a matrix $\mathcal{P}^{(i)}=\rm{diag}(1,0,0)$ in the respective basis $(l_i,l_i^{\perp^{(1)}},l_i^{\perp^{(2)}})$ where $i=1,2,3$, called the projection matrix. Since the projection matrices are defined in the RHN flavor basis, we can use Eq.(\ref{eq:leptonbasis}) to define a similarity transformation to represent the projection matrix in the charged flavor basis or in a different RHN flavor basis, which is shown as 
\begin{eqnarray}
    \label{eq:changeofbasis}
    \mathcal{P}^{(I)}_{i_Ij_I}=R^{(I)^\dagger}_{i_I \alpha} R^{(J)}_{\beta i_J}\mathcal{P}^{(J)}_{i_J j_J}R^{(J)^\dagger}_{j_J\beta}R^{(I)}_{\beta j_{I}}
\end{eqnarray}
where $I,J$ represents the RHN generation ($1,2,3$), $\alpha,\beta$ denotes the charged flavors ($e,\mu,\tau$) and $i_I,j_I$ represents the element location in the matrix corresponding to the $I^{\rm th}$ RHN generation. Using Eq.(\ref{eq:leptonbasis}), the matrix $R^{I}$
is given by
\begin{eqnarray}
    \label{eq:flavor_rotation}
    R^{I}
    =
    \begin{pmatrix}
        c_{I\tau}&\frac{1-|c_{I\tau}|^2}{\sqrt{|c_{Ie}|^2+|c_{I\mu}|^2}} & 0\\
        c_{I\mu}&\frac{-c_{I\mu} c_{I_\tau}^*}{\sqrt{|c_{Ie}|^2+|c_{I\mu}|^2}}& \frac{c_{Ie}^*}{\sqrt{|c_{Ie}|^2+|c_{I\mu}|^2}}\\
        c_{Ie}&\frac{-c_{Ie}c_{I\tau}^*}{\sqrt{|c_{Ie}|^2+|c_{I\mu}|^2}}&\frac{-c_{I\mu}^*}{\sqrt{|c_{Ie}|^2+|c_{I\mu}|^2}}
    \end{pmatrix}
\end{eqnarray}
Now, using Eq.(\ref{eq:changeofbasis}), and \ref{eq:flavor_rotation}, we can define all the projection matrices $\mathcal{P}^{(I)}$ in the RHN flavor basis $(l_1,l_1^{\perp^{(1)}},l_1^{\perp^{(2)}})$. In this basis, the density matrix equations are given as
\begin{eqnarray}
    \label{eq:Densitymatrixeq}
     \frac{d N_{N_1}}{dz} &=& - (D_1+S_1)(N_{N_1}-N_{N_1}^{eq}),\nonumber\\
        \frac{d N_{N_2}}{dz} &=& -  (D_2+S_2)(N_{N_2}-N_{N_2}^{eq}),\nonumber\\
        \frac{d N_{N_3}}{dz} &=& - (D_3+S_3)(N_{N_3}-N_{N_3}^{eq}),\nonumber\\
        \frac{dN^{B-L}_{ij}}{dz}&=&\sum_{k=1}^3\bigg(-\varepsilon^{(k)}_{ij}D_k(N_{N_k}-N_{N_k}^{eq})-\nonumber\\&&\frac{1}{2}W_k\{\mathcal{P}^{(k)},N^{B-L}\}_{ij}\bigg)
\end{eqnarray}
where, $z=M_1/T$, $\varepsilon^{(k)}_{ij}=\varepsilon_k \mathcal{P}^{(k)}_{ij}$ is the CP asymmetry matrix, $N^{B-L}_{ij}$ is the $B-L$ asymmetry matrix, $k$ defines the generation of RHNs($1,2,3$). Here, $D_k(z)=\frac{z\Gamma_{D_k}(\frac{M_k}{M_1}z)}{\mathcal{H}(M_k)}$ which accounts for the decays of $N_k$. $S_k$, $k=1,2,3$ constitutes $\Delta L=1$ scattering processes involving $N_k$. $W_k$ represents the washout term due to inverse decays, $\Delta L=1$ and $\Delta L =2$ scatterings.  The Hubble expansion rate is given by
 \begin{equation}
\mathcal{H} \approx \sqrt{\frac{8\pi^3g_*}{90}}\frac{T^2}{M_{\rm Pl}}\approx1.66\sqrt{g_*}\frac{T^2}{M_{\rm Pl}},    
 \end{equation}
where, $g_*=g_{\rm SM}=106.75$ is the total number of degrees of freedom, $M_{\rm Pl}=1.22 \times 10^{19} ~\rm GeV$ is the Planck mass, and $T$ is the temperature of the thermal bath.

The decay and scattering terms $D$ and $S$ depend on the \textit{effective neutrino mass} defined as
\begin{eqnarray}
    \tilde{m}_i = \frac{(m_D^\dagger m_D)_{ii}}{M_i},
\end{eqnarray}
which has to be compared with the equilibrium neutrino mass
\begin{eqnarray}
    m_* = \frac{16 \pi^{5/2}\sqrt{g_*}}{3\sqrt{5}}\frac{v^2}{M_{\rm Pl}}\approx 1.08 \times10^{-3} \rm{eV}.
\end{eqnarray}
Then we define the washout parameter as
\begin{eqnarray} \label{eq:washoutparam}K_i=\frac{\Gamma_i(z=\infty)}{\mathcal{H}(z=1)}=\frac{\tilde{m}_i}{m_*}.
\end{eqnarray}
This quantity defines whether or not the decay of $N_i$ is in equilibrium.

The $B-L$ asymmetry obtained in Eq.(\ref{eq:Densitymatrixeq}) is then transferred to the baryon asymmetry via EW sphaleron process and can be observed as:
\begin{eqnarray}
\label{eq:etaB}
    \eta_B = \frac{a_{\rm sph}}{f}N_{B-L}^{\rm final}
\end{eqnarray}
where $a_{\rm sph}=28/79$ is the fraction of $B-L$ asymmetry transferred to baryon asymmetry and $f=N_{\gamma}^{\rm rec}/N_{\gamma}^*=2387/86$ is the dilution factor. Therefore, the correct $B-L$ asymmetry, considering the central value of the observed baryon asymmetry $\eta_B^{\rm obs}=6.14\times10^{-10}$ is $N_{B-L}^{\rm obs}\simeq4.81\times10^{-8}$.
If we can consider the three ``flavor vectors" to be almost parallel to each other ($l_1||l_2||l_3$), then we can describe the dynamics by simple Boltzmann equations instead of the density matrix given by
\begin{eqnarray}\label{eq:BE}
        \frac{d N_{N_1}}{dz} &=& - (D_1+S_1)(N_{N_1}-N_{N_1}^{eq}),\nonumber\\
        \frac{d N_{N_2}}{dz} &=& -  (D_2+S_2)(N_{N_2}-N_{N_2}^{eq}),\nonumber\\
        \frac{d N_{N_3}}{dz} &=& - (D_3+S_3)(N_{N_3}-N_{N_3}^{eq}),\nonumber\\
        \frac{d N_{B-L}}{dz} &=& - \sum_{i=k}^3 \big(\varepsilon_k D_k(z) (N_{N_k}-N_{N_k}^{eq}) +\nonumber\\&& W_{k}(z)N_{B-L}(z)\big),\label{eq:be}
\end{eqnarray}
We can easily get Eq.(\ref{eq:be}) from Eq.(\ref{eq:Densitymatrixeq}) by taking a trace of the whole $B-L$ equation in the condition ($l_1||l_2||l_3$). However, if all the ``flavor vectors" are parallel to each other, then this will imply that the coefficient matrix $C=\{c_{i\alpha}\}$ is rank 1, and since the coefficient $c_{i\alpha} =\frac{Y_{\alpha i}}{\sqrt{(Y^\dagger Y)_{ii}}}$ at tree level, this means that the Yukawa coupling matrix $Y$ is rank one and will lead to only single non-zero light neutrino mass obtained from seesaw mechanism. But the current neutrino oscillation experiment demands that at least two of the light neutrinos are massive. In order for us to have at least 2 non-zero neutrino masses, at most two out of the three ``flavor vectors" can be parallel to each other.

Now, to show the effects of heavy neutrino flavor projection on the final $B-L$ asymmetry, we shall consider two extreme situations.

\underline{Case-1:} when $l_1$ and $l_2$ are parallel to each other.

\underline{Case-2:} when $l_1$ and $l_2$ are orthogonal to each other.

\begin{figure}[h]
    \centering
    \includegraphics[scale=0.4]{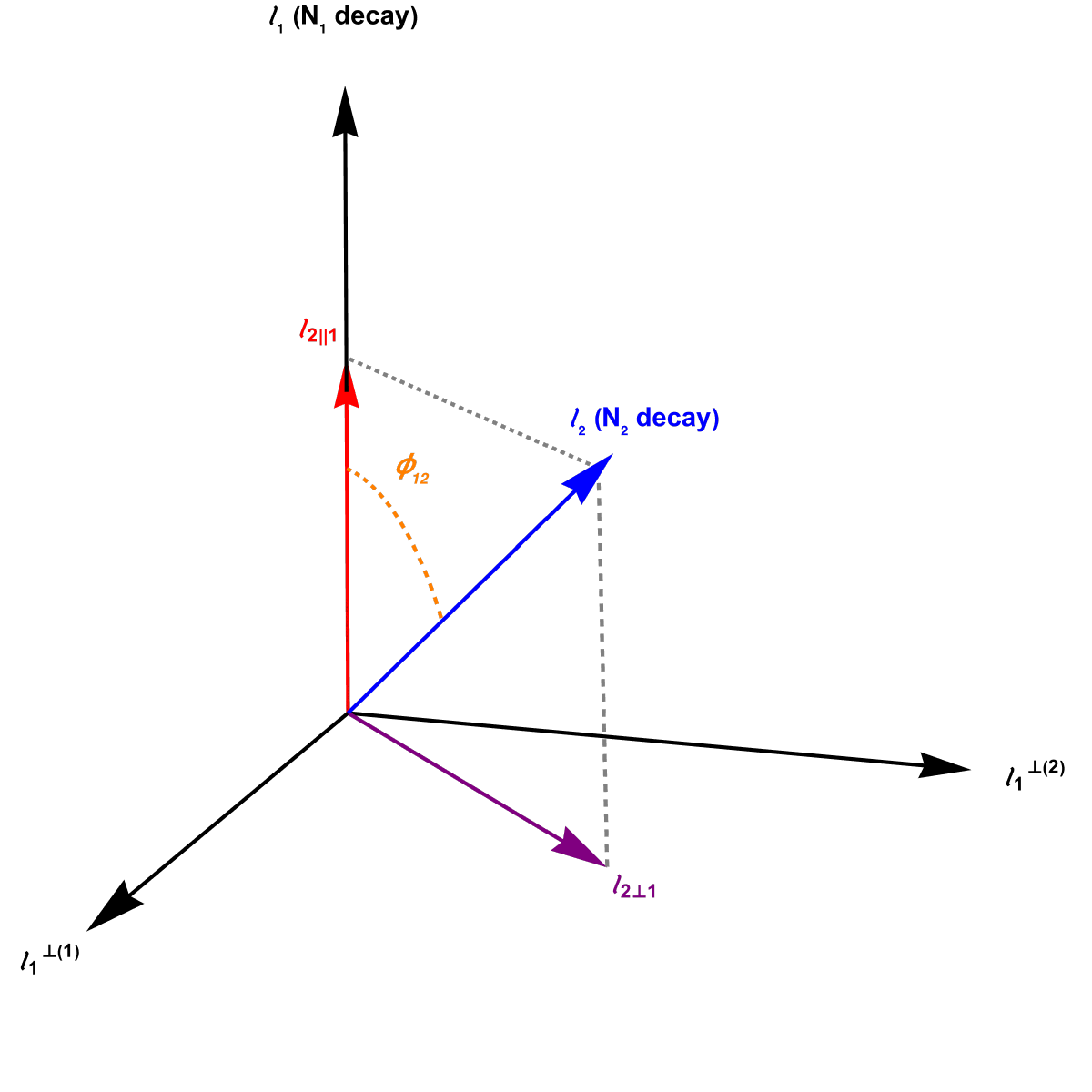}
    \caption{Figure illustrating the alignment of ``flavor vectors" in the $(l_1, l_1^{\perp (1)},l_1^{\perp (2)})$ basis in flavor space.}
    \label{fig:projectionplot}
\end{figure}

For simplicity, we consider only the effects of decay and inverse decay when solving the DMEs. To simplify the situation, we consider only $N_1$ and $N_2$, whose decay produces the lepton asymmetry. Now we consider $N_i$ decays to $l_i=\frac{Y_{\alpha i}}{\sqrt{(Y^\dagger Y)_{ii}}}l_\alpha$, where $i=1,2$ and $\alpha=e,\mu,\tau$.
To show this geometric interpretation, we define the angle between these two ``flavor vectors" in the flavor space as
\begin{eqnarray}
    \cos\phi_{ij}=\frac{\sum_{\alpha}Y_{\alpha i}Y^*_{\alpha j}}{\sqrt{(Y^\dagger Y)_{ii}}\sqrt{(Y^\dagger Y)_{jj}}}.
\end{eqnarray}
Fig.\ref{fig:projectionplot} shows a pictorial representation of the ``flavor vectors" $l_1$ and $l_2$ and the angle $\phi_{12}$ between them in the $(l_1, l_1^{\perp (1)},l_1^{\perp (2)})$ basis in flavor space.
Now for case-1, $\cos \phi_{12} \rightarrow1$ means that $l_1$ and $l_2$ are parallel to each other, and for case-2, $\cos\phi_{12}\rightarrow0$ means that $l_1$ is orthogonal to $l_2$. To illustrate the cases, we consider two benchmark points, BP1 and BP2, as mentioned in Table \ref{tab:benchmarks}. We then compute the projection matrices of the two RHNs for BP1 and BP2. The projection matrices $\mathcal{P}^{(1)}$ and $\mathcal{P}^{(2)}$ are given by Eq.(\ref{eq:parallelProjection}) and Eq.(\ref{eq:orthogonlProjection}). The ${11}$ component of the projection matrix $\mathcal{P}^{(2)}$ gives the component of $l_2$ parallel to $l_1$. For BP1, $\mathcal{P}^{(2)}_{11}=0.997$, which shows that $l_2$ is almost parallel to $l_1$. This belongs to our case-1. And for BP2, $\mathcal{P}^{(2)}_{11}=0.003$ which shows that $l_2$ is almost orthogonal to $l_1$. This belongs to our case-2.
\begin{table*}[t]
\centering
\setlength{\tabcolsep}{5pt}
\renewcommand{\arraystretch}{1.1}
\begin{tabular}{|l|c|c|c|c|c|c|c|c|c|}
\hline
 BPs& $x_1 (^\circ)$ & $y_1(^\circ)$ & $x_2(^\circ)$ & $y_2(^\circ)$ & $x_3(^\circ)$ & $y_3(^\circ)$ & $M_1$ (GeV) & $M_2$ (GeV) & $M_3$ (GeV) \\ 
\hline
BP1 & $-95.684$ & $-0.0503$ & $-402.789$ & $0.0627$ & $-0.3096$ & 
 $0.7575$ & $2.1\times10^{13}$ & $1.05\times10^{15}$ & $2.1\times10^{15}$ \\
\hline
BP2 & $-269.462$ & $-6.307$ & $19.882$ & $-0.01104$ & $-84.7951$ &$-92.705$ & $10^{12}$ & $5\times10^{13}$ & $10^{14}$ \\
\hline
\end{tabular}
\caption{Input parameters for the two benchmark points used in the plots.}
\label{tab:benchmarks}
\end{table*}
\begin{figure*}[tbh]
    \centering
    \includegraphics[scale=0.45]{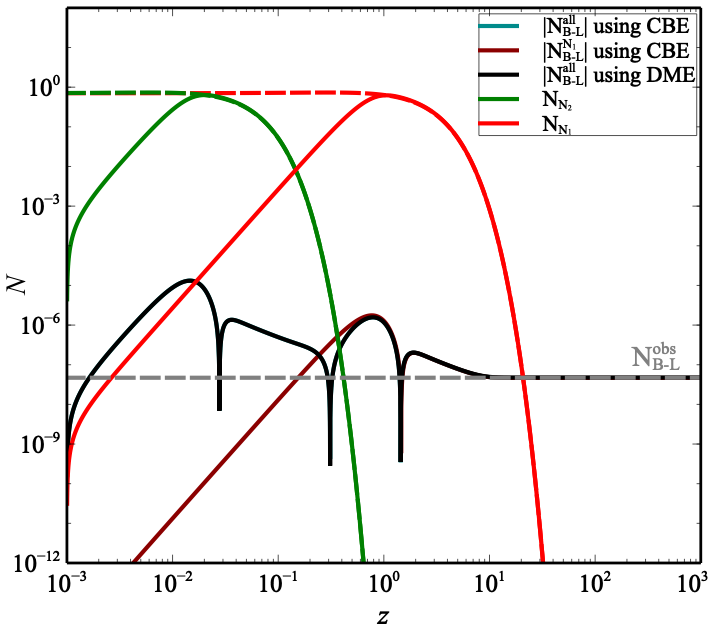}
    \includegraphics[scale=0.45]{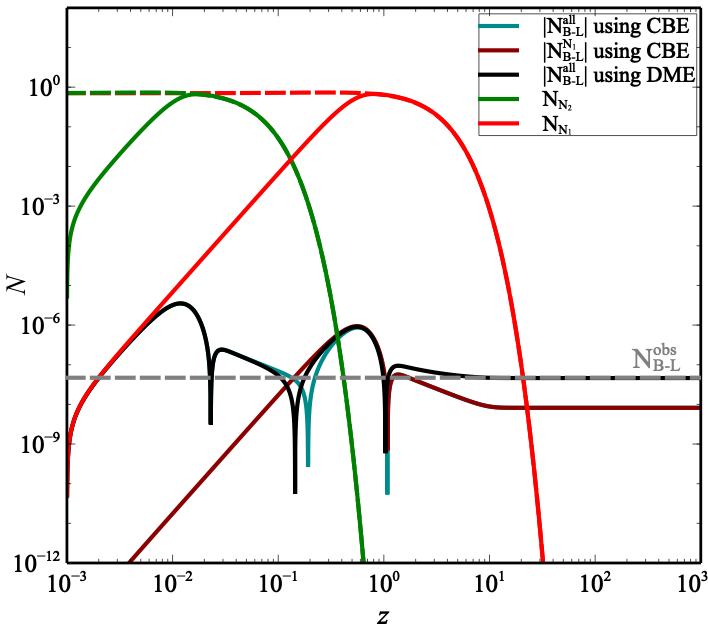}
    \caption{\textit{Left}: Figure illustrating the effect of ``flavor vector" in parallel orientation ($\cos\phi_{12}\rightarrow1$) (Yukawa coupling texture) on the final $B-L$ asymmetry. (BP-1: $K_{1}=21.62$ and $ K_{2}=25.06$). \textit{Right}: figure illustrating the effect of ``flavor vector" in orthogonal orientation ($\cos\phi_{12}\rightarrow0$) (Yukawa coupling texture) on the final $B-L$ asymmetry. (BP-2: $K_{1}=54.50$ and $ K_{2}=45.15$).}
    \label{fig:parallelbp}
\end{figure*}

In the \textit{left} panel of Fig.\ref{fig:parallelbp}, we show the evolution of the abundances of RHNs and the lepton asymmetries for the BP1 as given in Table \ref{tab:benchmarks}. The Yukawa coupling for this benchmark point is given as
\begin{widetext}
    \begin{eqnarray}
        \label{eq:BP1Yukawa}
        Y=
        \begin{pmatrix}
-1.04\times10^{-2} + 1.58\times10^{-2} i &
-7.31\times10^{-2} + 9.28\times10^{-2} i &
-1.05\times10^{-2} + 4.40\times10^{-1} i \\
-6.02\times10^{-4} - 8.61\times10^{-2} i &
2.55\times10^{-3} - 6.87\times10^{-1} i &
2.45\times10^{-2} + 4.12\times10^{-1} i \\
2.73\times10^{-4} - 9.10\times10^{-2} i &
-3.82\times10^{-3} - 6.66\times10^{-1} i &
2.52\times10^{-2} - 4.98\times10^{-1} i
\end{pmatrix}.
    \end{eqnarray}
\end{widetext}
    
Here, both RHNs $N_1$ and $N_2$ are in the strong washout regime. Here, the black curve denotes the $B-L$ asymmetry produced by both $N_1$ and $N_2$ using the DM formalism (\textit{i.e.,} Eq.(\ref{eq:Densitymatrixeq})), the dark-red curve denotes the $B-L$ asymmetry produced by the decay of $N_1$ alone, and the dark-cyan curve represents the $B-L$ asymmetry produced by both $N_1$ and $N_2$ using the Classical Boltzmann equation (CBE) (\textit{i.e.,} Eq.(\ref{eq:BE})). The red and green curves denote the comoving number densities of $N_1$ and $N_2$, respectively. The dashed red and blue curves denote the respective comoving equilibrium number densities. And finally the gray dashed line represents the correct $B-L$ asymmetry. In this plot, we observe that there is no difference in the final asymmetry produced by $N_1$ alone compared to that produced by both $N_1$ and $N_2$ using both approaches. Here, all the RHNs are in the strong washout regime, so this result is expected. This says that since the ``flavor vectors" are parallel to each other, any asymmetry produced by heavier RHN ($N_2$) is fully washed out by the inverse decays of the lighter RHN $N_1$. This also shows that when $l_1||l_2$, then the DME Eq.(\ref{eq:Densitymatrixeq}) collapses to the CBE Eq.(\ref{eq:BE})

On the other hand, in the \textit{right} panel of Fig.\ref{fig:parallelbp}, we show the evolution of the abundances and asymmetries for the BP2 as given in Table \ref{tab:benchmarks}. The Yukawa coupling matrix for BP2 is given as
\begin{widetext}
    \begin{eqnarray}
        \label{eq:BP2Yukawa}
        Y=
        \begin{pmatrix}
8.23\times10^{-4} - 2.10\times10^{-2} i &
2.33\times10^{-2} - 7.56\times10^{-2} i &
-2.11\times10^{-1} - 1.71\times10^{-2} i \\
1.04\times10^{-3} - 3.66\times10^{-2} i &
-2.79\times10^{-4} + 1.35\times10^{-1} i &
-2.87\times10^{-1} - 1.55\times10^{-2} i \\
-3.13\times10^{-3} + 1.15\times10^{-2} i &
-4.30\times10^{-3} + 2.35\times10^{-1} i &
1.73\times10^{-1} + 2.97\times10^{-2} i
\end{pmatrix}.
    \end{eqnarray}
\end{widetext}
In this case, all the RHNs are in the strong washout regime. The color code for the curves in the plot is same as the previous one. Here we observe that, within the classical Boltzmann treatment, the $B-L$ asymmetry obtained from $N_1$ alone is indistinguishable from that obtained when both $N_1$ and $N_2$ are included. However, within the DM framework, there is a clear difference in the final $B-L$ asymmetry produced by only $N_1$ in comparison to the asymmetry produced by both $N_1$ and $N_2$. This shows that in this case, there is some memory left of the asymmetry produced by the heavier RHN $N_2$. The physical essence of this benchmark point is that the “flavor vectors” are almost orthogonal to each other, which means that the asymmetry produced by $N_2$ is almost fully protected from the inverse-decay washout effects of $N_1$, and therefore the final asymmetry remaining is basically the sum of the asymmetries produced by $N_1$ and $N_2$. This essence is captured by the density-matrix formalism and therefore differs from the classical Boltzmann approach. Therefore, we see that this case goes against the conventional assumption that the lightest RHN fully washes out any asymmetry previously produced, and thus shows the effect of the Yukawa coupling texture (projection effect) on the final $B-L$ asymmetry.

\subsection{Washout Regime Analysis}\label{sec:washoutregimeana}

As previously defined in Eq.(\ref{eq:washoutparam}) we have
\begin{eqnarray}
    K_i =\frac{\tilde{m}_iM_i^2/8\pi v^2}{1.66\sqrt{g_*}M_i^2/M_{\rm Pl}}=\frac{\tilde{m}_i}{1.0697\times 10^{-3} ~\rm eV},
\end{eqnarray}
where $\tilde{m}_i=\frac{v^2(Y_{\nu}^\dagger Y_{\nu})_{ii}}{2M_i}$. If $K_i<1$ ($\tilde{m}_i<1.0697\times 10^{-3} ~\rm eV$), then the $i^{\rm th}$ RHN is in the weak washout regime. On the other hand, if $K_i>1$ ($\tilde{m}_i>1.0697\times 10^{-3} ~\rm eV$) then the $i^{\rm th}$ RHN is in the strong washout regime. Using Eq.(\ref{eq:casasiba}) we can write
\begin{eqnarray}\label{eq:mtilde}
\tilde{m}_i&=&\frac{1}{M_i}[(U^* D_{\sqrt{m}} R^T D_{\sqrt{M}})^\dagger(U^* D_{\sqrt{m}} R^T D_{\sqrt{M}})]_{ii}
    \nonumber,\\&=&\frac{1}{M_i}[(D_{\sqrt{M}}R^* D_{\sqrt{m}}U^T)(U^* D_{\sqrt{m}} R^T D_{\sqrt{M}})]_{ii}\nonumber,\\&=&\frac{1}{M_i}[D_{\sqrt{M}}R^* D_m R^T D_{\sqrt{M}}]_{ii}. 
\end{eqnarray}
By using the generalized rotation matrix, $R$ given by Eq.(\ref{eq:Rmatrix}) and putting it in the above Eq.(\ref{eq:mtilde}) we get
\begin{eqnarray}\label{eq:effectivem}
\tilde{m}_1&=&|\cos \theta_1 \cos \theta_2|^2 m_1 +|\cos \theta_1 \sin \theta_2 \sin \theta_3  \nonumber\\&&-\sin \theta_1 \cos \theta_3|^2 m_2 + |\cos \theta_1 \sin \theta_2 \cos \theta_3\nonumber\\&&+ \sin \theta_1 \sin \theta_3|^2 m_3,\nonumber\\
\tilde{m}_2&=&|\sin \theta_1 \cos \theta_2|^2 m_1 + |\sin \theta_1 \sin \theta_2 \sin \theta_3\nonumber\\&& + \cos \theta_1 \cos \theta_3|^2 m_2 + |\sin \theta_1 \sin \theta_2 \cos \theta_3 \nonumber\\&&- \cos \theta_2 \cos \theta_3|^2 m_3,\nonumber\\
\tilde{m}_3&=&|\sin \theta_2|^2 m_1 +|\cos \theta_2 \sin \theta_3|^2 m_2 \nonumber\\&&+ |\cos \theta_2 \cos \theta_3|^2 m_3.
\end{eqnarray}

So from the above equation we notice that condition for washout regime, either $K_i<1$ ($\tilde{m}_i<1.0697\times 10^{-3} ~\rm eV$) or $K_i>1$ ($\tilde{m}_i>1.0697\times 10^{-3} ~\rm eV$) depends on the absolute mass of the light neutrinos which can be obtained from the neutrino oscillation data. Now we ask the question, for a given set of light neutrino masses ($m_1, m_2, m_3$), how many RHNs can be in the weak washout regime ($\tilde{m_i}<1.0697\times 10^{-3} ~\rm eV \,\,\forall \,i$). This question can be answered from a simple analysis using the method of contradiction as given below. For normal mass ordering (NO) of the light active neutrinos, we set $m_1\rightarrow0$ for simplicity, and then $m_2=\sqrt{|\Delta m_{21}^2|}$ and $m_3=\sqrt{\Delta m_{31}^2}$.
\begin{figure*}[tbh]
\centering
 \includegraphics[scale=0.42]{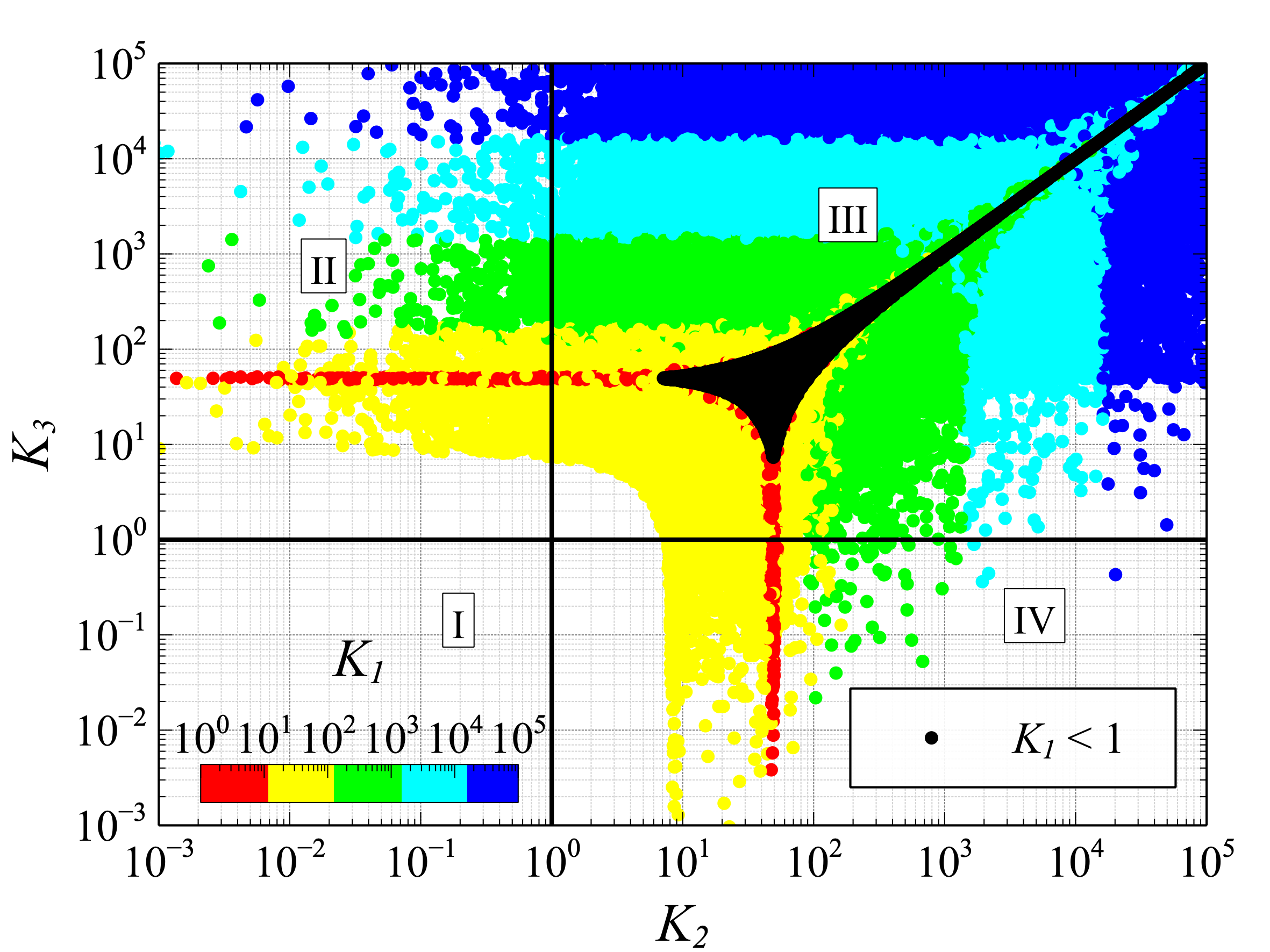}
  \includegraphics[scale=0.42]{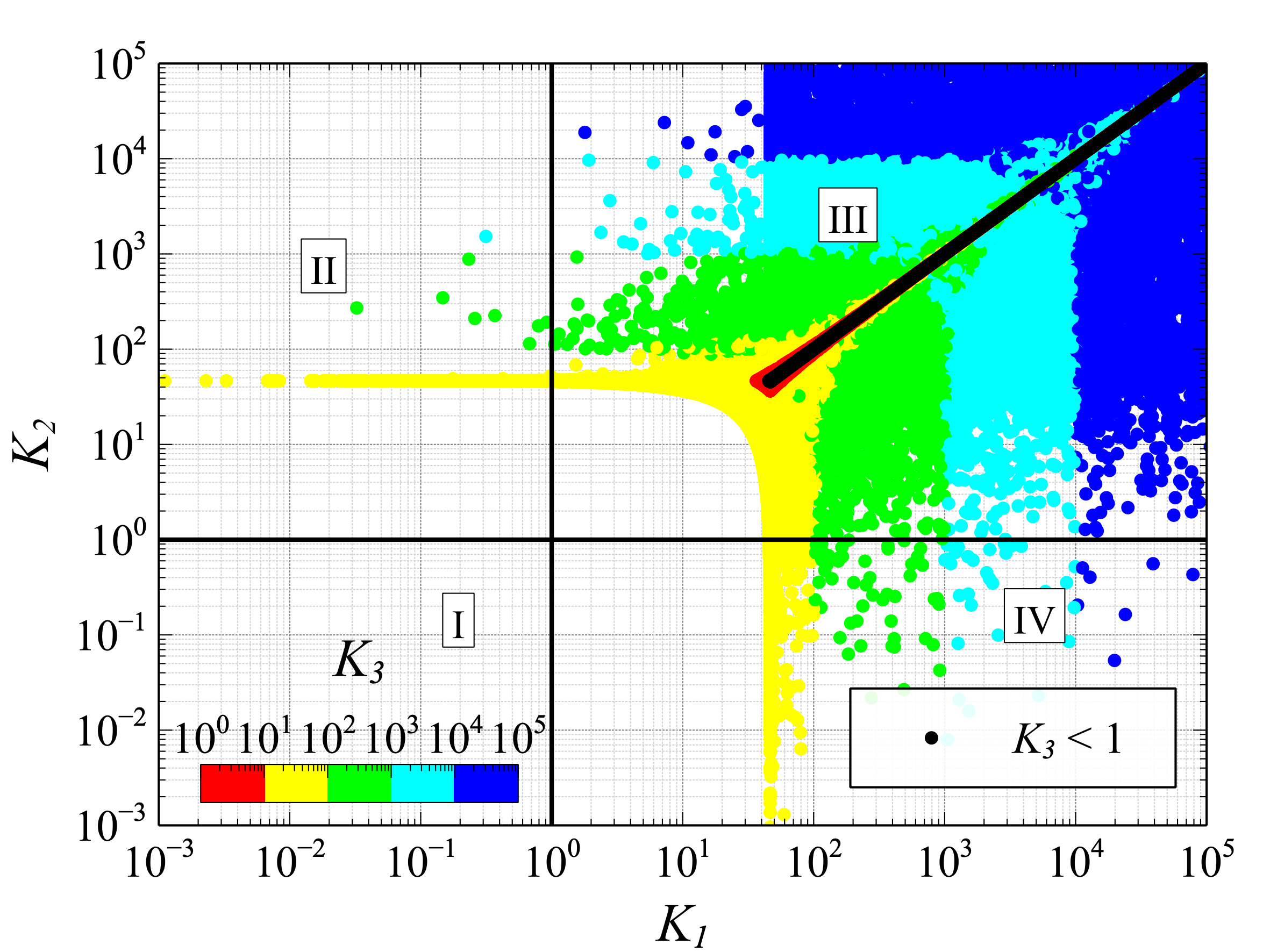}
  \caption{Plot illustrating the constraints between $K_1$, $K_2$, and $K_3$ and thereby showing that at most only one of the three $RHNs$ can be in weak washout regime in the case of normal ordering (\textit{left}), and inverse ordering (\textit{right}).}
  \label{fig:washoutregime}
\end{figure*}
Let us assume all RHNs to be in weak washout regime ($\tilde{m_i}<1.0697\times 10^{-3} ~\rm eV \,\,\forall\,i$). Now assuming $\theta_2 = \theta_3 = \frac{\pi}{2}$ in Eq.(\ref{eq:effectivem}) we get $\tilde{m_3}\longrightarrow0$ and
\begin{equation}
    \begin{split}
        &|\cos \theta_1|^2 m_2 + |\sin \theta_1|^2 m_3 < 1.0697 \times 10^{-3} ~\rm eV,\\
        &|\sin \theta_1|^2 m_2 + |\cos \theta_1|^2 m_3 < 1.0697 \times 10^{-3} ~\rm eV.
    \end{split}
\end{equation}
From the above inequalities, we get
\begin{equation}\label{eq:NO}
    m_2+m_3 < 2.1394 \times 10^{-3}~\rm  eV.
\end{equation}

Similarly, in the case of inverse mass ordering (IO) of the light active neutrinos, we set $m_3\rightarrow0$, then $m_1=\sqrt{|\Delta m_{31}^2|}$ and $m_2=\sqrt{|\Delta m_{31}^2|+\Delta m_{21}^2}$. Now if we take all the RHNs to be in weak washout regime ($\tilde{m_i}<1.0697\times 10^{-3} ~{\rm eV} \,\,\forall\,i$) and $\theta_2 =\theta_3 = \frac{\pi}{2}$, we get $\tilde{m_3}\rightarrow0$ and
\begin{equation}
    \begin{split}
        &|\cos \theta_1|^2 m_1 + |\sin \theta_1|^2 m_2 < 1.0697 \times 10^{-3} ~\rm eV,\\
        &|\sin \theta_1|^2 m_1 + |\cos \theta_1|^2 m_2 < 1.0697 \times 10^{-3}~\rm  eV,
    \end{split}
\end{equation}
which gives,
\begin{equation}\label{eq:IO}
    m_1+m_2 < 2.1394 \times 10^{-3}~\rm  eV.
\end{equation}
From Eqs.\ref{eq:NO} and \ref{eq:IO}, we see that the inequalities cannot be satisfied by the low-energy neutrino oscillation data\cite{deSalas:2020pgw,ParticleDataGroup:2024cfk}. This implies that our initial assumption that all three RHNs are simultaneously in the weak washout regime is incorrect. On the other hand, from Eq.(\ref{eq:NO}) we can see that the condition $m_2 +m_3 > 2.1394\times 10^{-3} ~\rm eV$ ($m_1+m_2 > 2.1394 \times 10^{-3}~\rm  eV$) can be satisfied by the low-energy neutrino oscillation data for NO (IO). So the above analysis shows that when $N_3$ is in the weak washout regime, both $N_1$ and $N_2$ must be in the strong washout regime. This same exercise can be repeated by setting $\tilde{m}_1,\tilde{m}_2 \rightarrow0$, which shows that for a given set of $m_1,m_2,m_3$ at most one of the RHNs can be in the weak washout regime\cite{DiBari:2005st}. 

Although in the above analysis we fixed the value of $\theta_{2},\theta_3=\pi/2$, it can be generalized by considering arbitrary complex values of $\theta_i \, \forall \, i$ as shown in Fig.\ref{fig:washoutregime}. For simplicity, we choose the lightest neutrino mass $m_1$ ($m_3$) to be zero for NO (IO).
We illustrate the results, (i) for normal ordering in the plane of $K_3-K_2$ with $K_1$ in color code as shown in Fig.\ref{fig:washoutregime} (\textit{left}) and (ii) for inverse ordering in the plane of $K_2 - K_1$ with $K_3$ in the color code as shown in Fig.\ref{fig:washoutregime} (\textit{right}). For analysis purpose, we divide Fig.\ref{fig:washoutregime} (\textit{left}) into four zones, (I) $K_2,K_3<1$, (II) $K_2<1, K_3>1$, (III) $K_2,K_3>1$ and (IV) $K_2>1,K_3<1$. In the zone (I), no points indicate that $N_2$ and $N_3$  simultaneously being in the weak washout regime is not possible. In zone (II), when $N_2$ is in the weak washout regime, $N_1$ and $N_3$ are in the strong washout regime. In zone (III), when $N_2$ and $N_3$ are in a strong washout regime, then we have two possibilities: either $N_1$ is in a strong washout regime (colored points), or it is in a weak washout regime (black points). Finally, in zone (IV), when $N_3$ is in the weak washout regime, then $N_1$ and $N_2$ are in the strong washout regime. The same analysis is valid for inverse ordering as shown in the Fig.\ref{fig:washoutregime}(\textit{right}). This validates the analytical proof we showed above. 

\section{Dynamical Evolution of $B-L$ asymmetry in thermal leptogenesis}\label{sec:result}

\begin{table*}[t]
\centering
\setlength{\tabcolsep}{5pt}
\renewcommand{\arraystretch}{1.1}
\begin{tabular}{|l|c|c|c|c|c|c|c|}
\hline
Cases & $x_1(^\circ)$ & $y_1(^\circ)$ & $x_2(^\circ)$ & $y_2(^\circ)$ & $x_3(^\circ)$ & $y_3(^\circ)$ & $M_1$ (GeV) \\ 
\hline
Case-1 (a) & $-374.803$ & $-34.2302$ & $250.481$  & $-8.19379$ & $4.42873$  & $2.18086$ & $10^{12}$ \\
 \hline
 Case-1 (b) & $42.079$ & $-7.9681$ & $-3.818\times 10^{-2}$  & $3.877\times 10^{-2}$ & $-443.96$  & $1.9254\times 10^{-3}$ & $10^{12}$ \\
 \hline
Case-2 & $-44.898$ & $-4.7806$ & $-8.43\times10^{-4} $& $1.66\times10^{-1}$ & $-8.23\times 10^{-2}$ &$5.71\times10^{-1}$ & $10^{12}$ \\
\hline
Case-3 & $-4.62\times 10^{-2}$ & $1.35 \times 10^{-3}$ & $448.112$ & $-13.045 $ & $274.443$ &$-1.78\times10^{-2}$& $10^{12}$ \\
\hline
Case-4 & $-95.632$ & $-2.616 \times 10^{-1}$ & $8.097\times 10^{-3}$ & $1.1025$ & $280.247$ &$6.2784$ & $3\times10^{12}$ \\
\hline
\end{tabular}
\caption{Input parameters for the four washout regimes.}
\label{tab:cases}
\end{table*}

\begin{figure*}[!]
    \centering
    \includegraphics[scale=0.42]{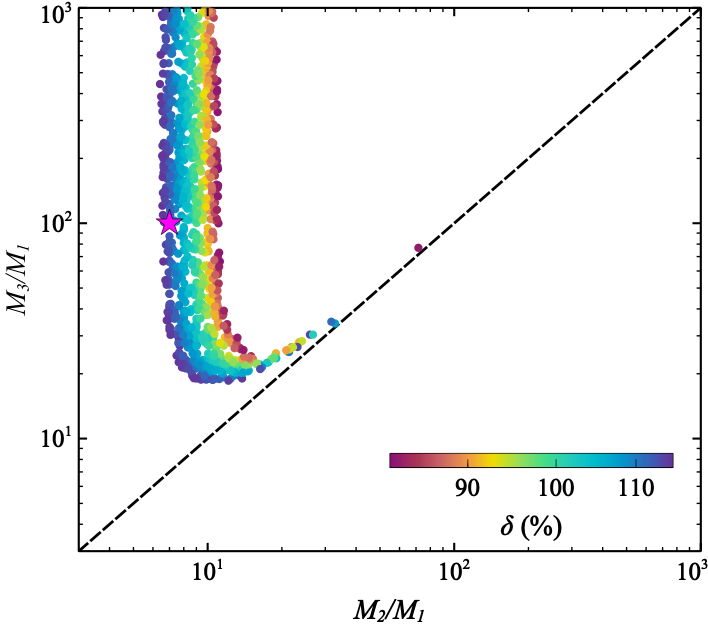} \includegraphics[scale=0.42]{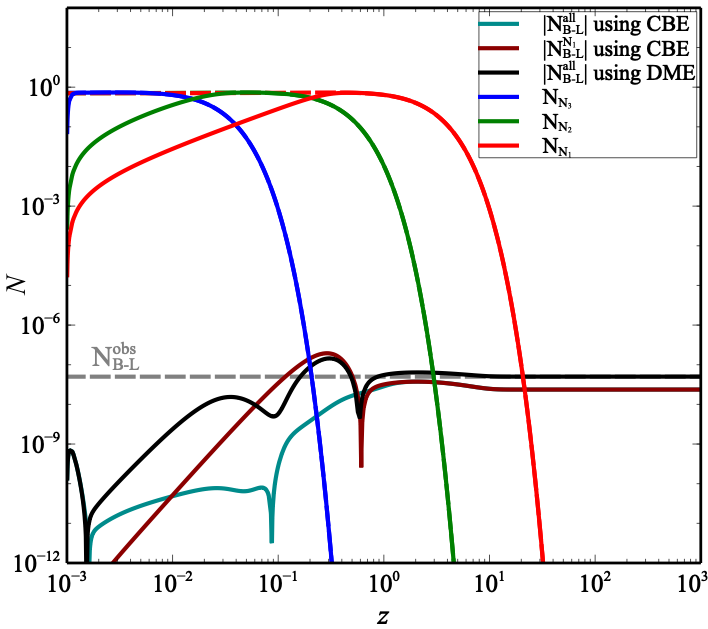} 
    \caption{[\textit{Left}:] Allowed parameter space in the plane $M_3/M_1$ vs $M_2/M_1$ when all the RHNs are in strong washout regime. The color code depicts the $\delta$ values representing the \textit{memory} effect. [\textit{Right}:] Cosmological evolution of comoving number densities of RHNs and $B-L$ asymmetries for the magenta star point when all RHNs are in the strong washout regime.}    \label{fig:Allstrongscattering}
\end{figure*}

\begin{figure*}[!]
    \centering
    \includegraphics[scale=0.42]{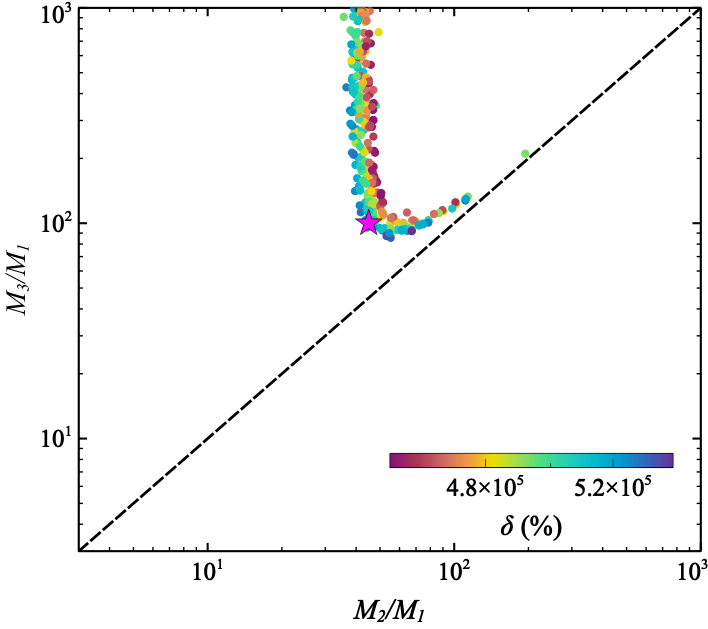} \includegraphics[scale=0.42]{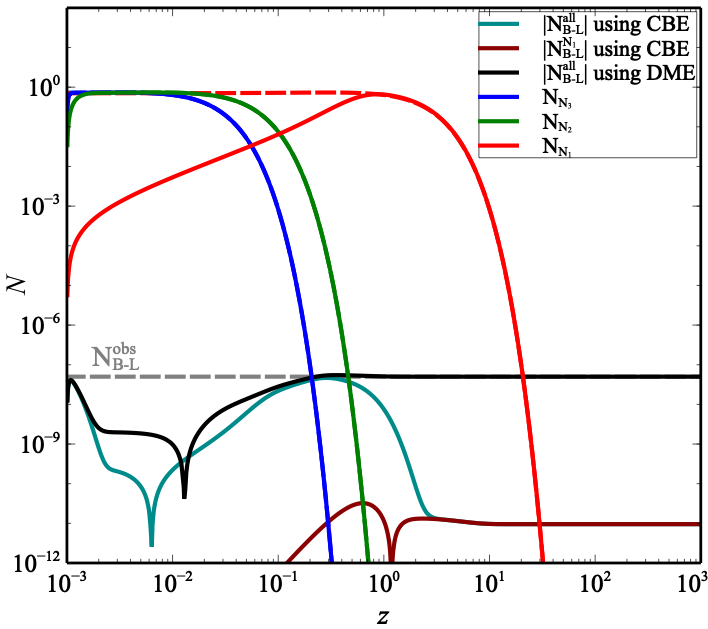} 
    \caption{[\textit{Left}:] Allowed parameter space in the plane $M_3/M_1$ vs $M_2/M_1$ when all the RHNs are in strong washout regime. The color code depicts the $\delta$ values representing the \textit{memory} effect. [\textit{Right}:] Cosmological evolution of comoving number densities of RHNs and $B-L$ asymmetries for the magenta star point when all RHNs are in the strong washout regime.}
    \label{fig:Allstrongscattering2}
\end{figure*}

Based on the discussion in section \ref{sec:washoutregimeana}, we concluded that at most one of the RHNs can be in a weak wash-out regime. This divides our analysis into four cases: (1) all three RHNs are in a strong washout regime, (2) $N_1$ is in a weak washout regime, (3) $N_2$ is in a weak washout regime, and (4) $N_3$ is in a weak washout regime.\\

As discussed in the literature \cite{Fukugita:1986hr,PhysRevD.45.455,PhysRevD.46.5331,Flanz:1994yx,Davidson:2008bu,Buchmuller:2004nz, Barbieri:1999ma}, in the type-I seesaw framework, $N_1$ is the only d.o.f. which is responsible for thermal leptogenesis. This is because any asymmetry produced by $N_{2,3}$ is presumed to be washed out by the lepton number-violating interactions of $ N_1$. However, when we take projection effects into account ,we saw that the asymmetry produced by $N_{2,3}$ may not be washed out fully even if $N_1$ is in the strong washout regime. To capture this effect, we define the parameter $\delta$ as
\begin{equation}\label{eq:delta}
    \delta=\frac{N_{B-L}^{{\rm all}}-N_{B-L}^1}{N_{B-L}^{{\rm1}}}\times100\%.
\end{equation}
where, $N^1_{B-L}$ denotes the final $B-L$ asymmetry when only $N_1$ contribution to $B-L$ asymmetry is under consideration, and $N^{{\rm all}}_{B-L}$ denotes the final $B-L$ asymmetry when all $N_1, N_2, N_3$ simultaneously contribute.
From Eq.(\ref{eq:delta}), $\delta\rightarrow0$ implies that solving the density matrix equation \ref{eq:Densitymatrixeq} considering all the RHNs is equivalent to solving Eq.(\ref{eq:BE}) by considering only $N_1$ contribution. In other words, $\delta\rightarrow0$ implies $N_1$ is the only d.o.f, and one can neglect the contribution from $N_{2,3}$ safely. Any non-zero values of $\delta$ imply a \textit{memory} effect on the final $B-L$ asymmetry. 

To generate the $B-L$ asymmetry dynamically, we solve the DMEs (Eq.(\ref{eq:Densitymatrixeq})), considering decay, inverse decay, and scatterings. It is worth noting that scatterings are dominant in the early times, while decay dominates in late times. We analyze each washout scenario individually in the plane of mass ratios $M_2/M_1$ and $M_3/M_1$, while keeping the CI parameters fixed as given in tab.\ref{tab:cases}. We focus only on the region of parameter space where the density matrix equation (DME) yields correct $N_{B-L}$, and corresponding to the observed $\eta_B$. In our numerical setup, we record all the points lying in the range: $5.1\times10^{-8}\gtrsim N_{B-L}\gtrsim4.3\times10^{-8}$. With this in mind, we analyze all the above-mentioned cases.
\begin{figure*}[!]
    \centering
    \includegraphics[scale=0.42]{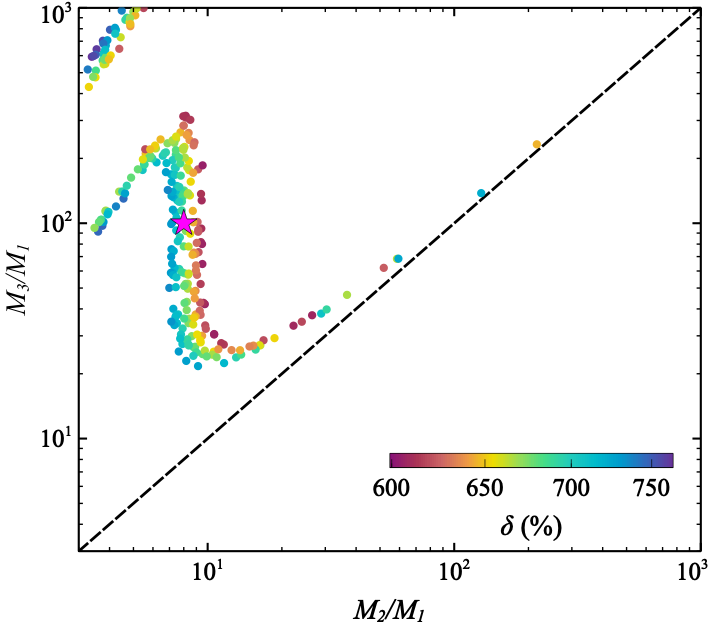}
    \includegraphics[scale=0.42]{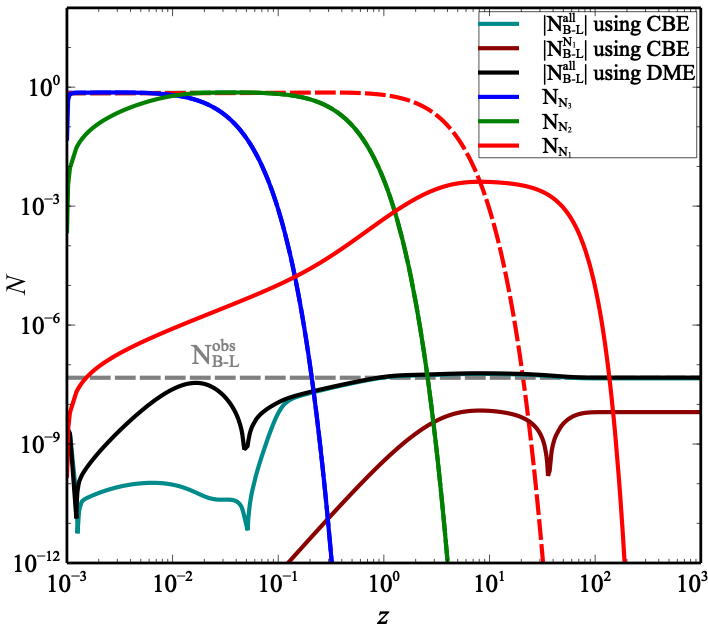}
    \caption{[\textit{Left}:] Allowed parameter space in the plane $M_3/M_1$ vs $M_2/M_1$ when $N_1$ is in weak washout regime. The color code depicts the $\delta$ values representing the \textit{memory} effect. [\textit{Right}:] Cosmological evolution of comoving number densities of RHNs and $B-L$ asymmetries for the magenta star point when $N_1$ is in the weak washout regime.}
    \label{fig:N1weakscattering}
\end{figure*}

\subsubsection{All RHNs are in strong washout regime}

In this case, we have taken all the RHNs to be in the strong washout regime by fixing the C.I. angles as given in the Table \ref{tab:cases} (case-1(a)). We vary the two mass ratios $M_2/M_1$ and $M_3/M_1$ and check the memory effect. The left panel of Fig.\ref{fig:Allstrongscattering} shows the values of $\delta$ (colored points) in the plane of $M_3/M_1$ and $M_2/M_1$ for a typical values of the washout parameters $K_1 = 42.60,\, K_2 = 13.77, \,K_3 =12.81$. And we have also fixed the mass of $N_1$ as $M_1=10^{12} \rm{GeV}$. Fixing the Casas--Ibarra parameters also fixes the projection matrices $(\mathcal{P}^{(1)},\, \mathcal{P}^{(2)},\, \mathcal{P}^{(3)})$ in the $(\ell_1,\, \ell_{1}^{\perp (1)},\, \ell_{1}^{\perp (2)})$ basis, as given in Eq.(\ref{eq:allstrongProjection}).\\
From Eq.(\ref{eq:allstrongProjection}), we obtain $\mathcal{P}^{(2)}_{11}=0.20$ and $\mathcal{P}^{(3)}_{11}=0.63$. This implies that a significant fraction of the asymmetry produced from $N_2$ decay is protected from the $N_1$ washout, whereas the asymmetry generated from $N_3$ decay is more strongly affected by the $N_1$ washout. Consequently, $N_2$ decay is expected to provide the dominant contribution to the final $B-L$ asymmetry. 
Figure~\ref{fig:Allstrongscattering} (left) shows the parameter space that yield the correct $B-L$ asymmetry within the density matrix formalism. It also highlights a significant deviation due to the memory effect ($70-130\%$). Due to the mass hierarchy among the RHNs ($M_3 > M_2 > M_1$), all points appear above the diagonal dashed-black line. Since a large fraction of the asymmetry generated from $N_3$ decay is washed out by $N_1$, the allowed parameter space mainly depends on $M_2/M_1$. 
We observe that the memory effect decreases as $M_2/M_1$ increases. As a result, the region with $M_2/M_1 \lesssim 6$ leads to an overproduction of asymmetry, while the region with $M_2/M_1 \gtrsim 11$ becomes underabundant. However, this trend does not always hold. When $M_3 \gtrsim M_2$, the two heavier RHNs decay nearly simultaneously and generate a larger asymmetry. Consequently, even parameter points with $M_2/M_1 > 11$ can yield the correct $B-L$ asymmetry in this regime.

The right panel of Fig.~\ref{fig:Allstrongscattering} illustrates a benchmark point with mass ratios $M_2/M_1 = 7$ and $M_3/M_1 = 100$, which is shown as a magenta start in Fig.~\ref{fig:Allstrongscattering}(left). This benchmark corresponds to a typical hierarchical spectrum ({\it i.e.} $M_3 > M_2 > M_1$) with the Yukawa coupling matrix as:
\begin{widetext}
    \begin{eqnarray}
    \label{eq:allstrongYukawa}
    Y=
\begin{pmatrix}
(-2.59 \times 10^{-3} + 4.85 \times 10^{-3} i) & (8.95 \times 10^{-3} + 2.98 \times 10^{-2} i) & (-7.72 \times 10^{-2} + 5.44 \times 10^{-2} i) \\
(-1.08 \times 10^{-3} - 2.65 \times 10^{-2} i) & (-1.01 \times 10^{-2} + 3.44 \times 10^{-2} i) & (-4.06 \times 10^{-2} - 6.57 \times 10^{-2} i) \\
(-1.62 \times 10^{-3} - 2.77 \times 10^{-2} i) & (-2.27 \times 10^{-2} - 2.53 \times 10^{-2} i) & (8.72 \times 10^{-2} - 1.51 \times 10^{-1} i)
\end{pmatrix}
\end{eqnarray}
\end{widetext}

The evolution of the RHN abundances $N_1$, $N_2$, and $N_3$ is shown in solid red, green, and blue, respectively. The $B-L$ asymmetry generated by $N_1$ alone is indicated by the dark-red curve, while the solid black curve shows the asymmetry obtained by including all three RHNs within the density-matrix framework. For comparison, the dark-cyan curve corresponds to the classical Boltzmann treatment considering all three RHNs contributing to the final $B-L$ asymmetry without projection effects. The gray-dashed line represents the correct $B-L$ asymmetry. In this case, the large Yukawa couplings also result in rapid inverse decays and scatterings, which quickly wash out the asymmetries generated by $N_2$ and $N_3$. This is clearly visible in the classical Boltzmann solution shown by the dark-cyan curve. However, due to the projection effects, a large portion of the asymmetry produced by heavier RHNs is not subjected to the washout effects. This is clearly visible in the black curve (obtained by solving Eq.(\ref{eq:Densitymatrixeq})), which retains a large asymmetry, illustrating the sizable memory effect.

The parameter space we got in the Fig.\ref{fig:Allstrongscattering} (left) is not unique and is completely dependent on the Yukawa coupling matrix and thereby on the CI parameters chosen. In order to demonstrate this, we consider another example of all strong washout regime where the DME gives correct relic in a different parameter space. The CI parameter chosen for this example is given in Table \ref{tab:cases} (Case-1 (b)). Fig \ref{fig:Allstrongscattering2} (left) shows a new parameter space in the plane of $M_2/M_1$ and $M_3/M_1$ for correct $B-L$ asymmetry generated from density matrix approach with $K_1=8.00$, $K_2=22.01$, and $K_3=26.77$. Here, the parameter space for correct relic has been shifted to $48\gtrsim M_2/M_1\gtrsim38$ with very small dependence on $N_3$. An interesting thing to notice here is that the memory effect in this case is much larger compared to the previous case. The reason mainly lies in the projection effects. From Eq.(\ref{eq:allstrongProjection2}), we can see that $\mathcal{P}_{11}^{(2)}=0.003$ and $\mathcal{P}_{11}^{(3)}=0.002$, which means the decays of both the RHNs $N_2$ and $N_3$ are completely perpendicular to the $N_1$ decay and therefore almost all of the asymmetry produced by heavier generation is protected which gives such a large memory effect. In order to show this nature in detail, we have also shown a benchmark plot, Fig. \ref{fig:Allstrongscattering} (right) with the mass splitting $M_2/M_1=45$ and $M_3/M_1=100$ as shown by the magenta star in Fig. \ref{fig:Allstrongscattering} (left). The corresponding Yukawa coupling matrix is given as
\begin{widetext}
    \begin{eqnarray}
    \label{eq:allstrongYukawa2}
    Y=
    \begin{pmatrix}
(1.31 \times 10^{-6} - 9.22 \times 10^{-3} i) & (1.04 \times 10^{-2} - 2.07 \times 10^{-2} i) & (2.78 \times 10^{-2} - 4.21 \times 10^{-2} i) \\
(-5.23 \times 10^{-4} - 1.10 \times 10^{-2} i) & (1.87 \times 10^{-2} + 1.31 \times 10^{-1} i) & (-2.74 \times 10^{-2} + 1.99 \times 10^{-1} i) \\
(-5.56 \times 10^{-4} + 8.71 \times 10^{-3} i) & (2.12 \times 10^{-2} + 1.28 \times 10^{-1} i) & (-2.68 \times 10^{-2} + 2.26 \times 10^{-1} i)
\end{pmatrix}
\end{eqnarray}
\end{widetext}
The color code is same as the previous benchmark plot. Here, we see that the heavier generations produce very large asymmetries which is visible from both black and darkcyan curve but only when we consider the projection effects can we get the correct $B-L$ asymmetry even when $N_1$ is heavily underabundant.

\subsubsection{$N_1$ is in weak washout regime}

Now, we consider the scenario where \( N_1 \) is in the weak washout regime. As indicated in Fig. \ref{fig:washoutregime}, if \( N_1 \) is in the weak washout regime, then the other two RHNs, \( N_2 \) and \( N_3 \), must necessarily be in the strong washout regime. To do this, we fix the C.I. parameters as given in Table \ref{tab:cases} (case-2). This gives us the washout parameters as $K_1=1.22\times10^{-3},\, K_2=27.79,$ and $K_3=27.93$. We also fix the mass of $N_1$ as $M_1=10^{12}\rm{GeV}$. Since we fix the C.I. parameter, this fixes the projection matrices, and in this case, they are given as Eq.(\ref{eq:N1weakProjection}). From Eq.(\ref{eq:N1weakProjection}), we see that $\mathcal{P}^{(2)}_{11}=0.059$ and $\mathcal{P}^{(3)}_{11}=0.707$, meaning that the decays of $N_2$ is almost completely perpendicular but decays of $N_3$ are almost completely parallel to $N_1$ decays. That is the asymmetry produced by $N_{3}$ is largely affected by $N_1$ washout effects, but $N_2$ being almost perpendicular largely affects the final asymmetry. The resulting value of $\delta$ is shown in Fig.\ref{fig:N1weakscattering} (\textit{left}), in the plane of $M_3/M_1-M_2/M_1$.

As $N_1$ is in weak washout regime, the Yukawa coupling of \( N_1 \) is small. Thus, the pre-existing asymmetry left by $N_2$ and $N_3$ persist in the final result, and we therefore see a large \textit{memory} effect. Due to the mass hierarchy, all points lie above the diagonal dashed-black line. We see that as mass of $N_2$ increases memory effect decreases. Therefore the region $M_2/M_1<7$ is mostly overabundant, and region where $M_2/M_1>$10 is mostly underabundant.

Now to explain the scenario more clearly, we choose a benchmark point as shown with a star in the left panel of Fig.\ref{fig:N1weakscattering}. We show the evolution of the asymmetries in the right panel of Fig.\ref{fig:N1weakscattering} and the color code remains the same as in the right panel of Fig.\ref{fig:Allstrongscattering}. This BP corresponds to $M_2/M_1=9$ and $M_3/M_1$=100 with the Yukawa coupling matrix as
\begin{widetext}
    \begin{eqnarray}
    \label{eq:N1weakYukawa}
    Y=
        \begin{pmatrix}
(-7.76 \times 10^{-5} - 3.93 \times 10^{-6} i) & (-5.74 \times 10^{-3} + 2.84 \times 10^{-2} i) & (1.43 \times 10^{-2} + 2.73 \times 10^{-2} i) \\
(-1.91 \times 10^{-4} - 1.12 \times 10^{-5} i) & (7.48 \times 10^{-3} - 3.31 \times 10^{-2} i) & (1.33 \times 10^{-2} + 2.75 \times 10^{-1} i) \\
(1.64 \times 10^{-6} + 1.76 \times 10^{-5} i) & (4.52 \times 10^{-3} - 7.64 \times 10^{-2} i) & (2.63 \times 10^{-2} + 1.46 \times 10^{-1} i)
\end{pmatrix}.
    \end{eqnarray}
\end{widetext}

Here, the asymmetry is effectively produced by $N_2$ as any asymmetry produced by $N_3$ is washed out by $N_2$ and $N_1$ washout effects, and the asymmetry generated by $N_2$ survives with and without the projection effects as $N_1$ is in weak washout regime.

\begin{figure*}[!]
    \centering
\includegraphics[scale=0.42]{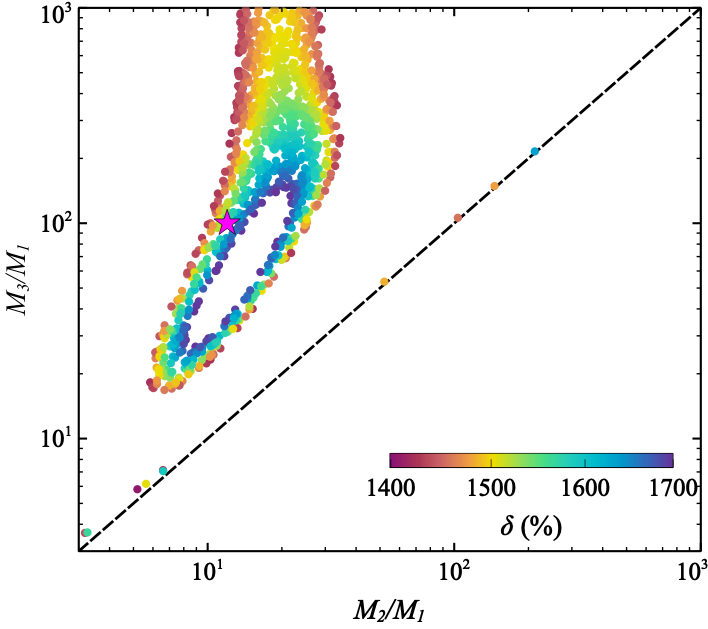}
\includegraphics[scale=0.42]{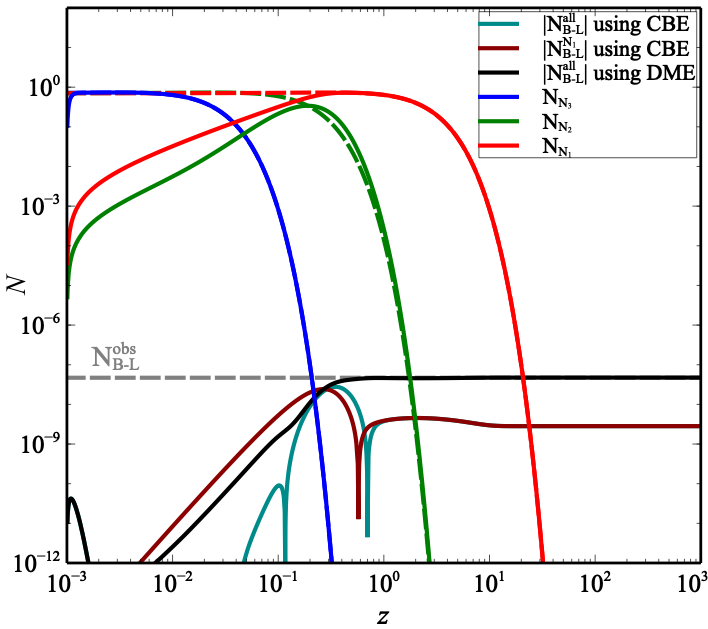}
    \caption{[\textit{Left}:] Allowed parameter space in the plane $M_3/M_1$ vs $M_2/M_1$ when $N_2$ is in weak washout regime. The color code depicts the $\delta$ values representing the \textit{memory} effect. [\textit{Right}:] Cosmological evolution of comoving number densities of RHNs and $B-L$ asymmetries for the magenta star point when $N_2$ is in the weak washout regime.}
    \label{fig:N2weakscattering}
\end{figure*}

\subsubsection{$N_2$ is in weak washout regime}

We now move to the case where \( N_2 \) is in the weak washout regime, while \( N_1 \) and \( N_3 \) remain in the strong washout regime. We consider a typical case by taking the C.I.\ parameters as given in Table~\ref{tab:cases} (case-3). This gives us the washout parameters \( K_1 = 39.712, \, K_2 = 4.75\times 10^{-2}, \, K_3 = 10.98 \), and the mass of \( N_1 \) is fixed as \( M_1 = 10^{12}~\rm GeV \). As we fix the C.I. parameters, the projection matrices also get fixed, which are given in this case by Eq.(\ref{eq:N2weakProjection}). From Eq.(\ref{eq:N2weakProjection}), we see that \( \mathcal{P}^{(2)}_{11} = 0.008 \) and \( \mathcal{P}^{(3)}_{11} = 0.310 \), meaning that a large amount of asymmetry produced by \( N_2 \) and \( N_3 \) is protected from the \( N_1 \) washout effects. Only about one-third of the asymmetry produced by \( N_3 \) is subjected to $N_1$ washout.

Figure~\ref{fig:N2weakscattering} (\textit{left}) shows the variation of the values of \( \delta \) in the plane of \( M_2/M_1 \) and \( M_3/M_1 \) corresponding to correct $B-L$ asymmetry points. From Fig.~\ref{fig:N2weakscattering} (\textit{left}), we see that almost all the points in the plot show a large memory effect. That is, the asymmetries produced by \( N_2 \) and \( N_3 \) contribute significantly to the final \( B-L \) asymmetry. 

In this case both $N_2$ and $N_3$ affects the final $B-L$ asymmetry. As the asymmetry produced by both the RHNs are almost perpendicular we can see that the region for correct $B-L$ asymmetry increases almost linearly with $N_2$ and $N_3$ mass. But for very large $N_3$ mass compared to $N_2$, its effect almost ceases. This is due to the $N_3$ asymmetry being washed out by scattering processes. Another point to note is that the parameter space shown in the Fig.  \ref{fig:N2weakscattering} (left), has two regions disconnected from each other. The points outside are underabundant regions whereas the small hole with no points are overabundant region. This does not apply to the points near the diagonal as here the mass of $N_2$ and $N_3$ are close and this causes increase in CP asymmetry (resonance region) and they produce correct asymmetry simultaneously.

To explain this more we have shown a case study in the right panel of Fig.~\ref{fig:N2weakscattering} by choosing a benchmark point (marked by the magenta star in the \textit{left} panel of Fig.~\ref{fig:N2weakscattering}). For this point, we take \( M_2/M_1 = 12 \) and \( M_3/M_1 = 100 \), and the color code is the same as in the previous plots with the Yukawa coupling matrix for this case as:
\begin{widetext}
    \begin{eqnarray}
        \label{eq:N2weakYukawa}
        Y=
        \begin{pmatrix}
(5.38 \times 10^{-3} - 5.60 \times 10^{-3} i) & (5.34 \times 10^{-6} + 2.46 \times 10^{-3} i) & (1.36 \times 10^{-2} + 1.01 \times 10^{-1} i) \\
(2.72 \times 10^{-3} + 2.78 \times 10^{-2} i) & (1.63 \times 10^{-4} + 2.93 \times 10^{-3} i) & (-5.73 \times 10^{-2} + 1.22 \times 10^{-1} i) \\
(-1.81 \times 10^{-3} + 3.02 \times 10^{-2} i) & (1.53 \times 10^{-4} - 2.33 \times 10^{-3} i) & (-6.24 \times 10^{-2} - 7.99 \times 10^{-2} i)
\end{pmatrix}.
    \end{eqnarray}
\end{widetext}
Here, the asymmetry produced by $N_3$ remains protected from $N_{1,2}$ washout effects and is further increased by the asymmetry produced by $N_2$. This results in a large memory effect and the final value lying in the range of correct $B-L$ asymmetry even when $N_1$ is underabundant.

\subsubsection{$N_3$ is in weak washout regime}

\begin{figure*}[htbp!]
    \centering
\includegraphics[scale=0.42]{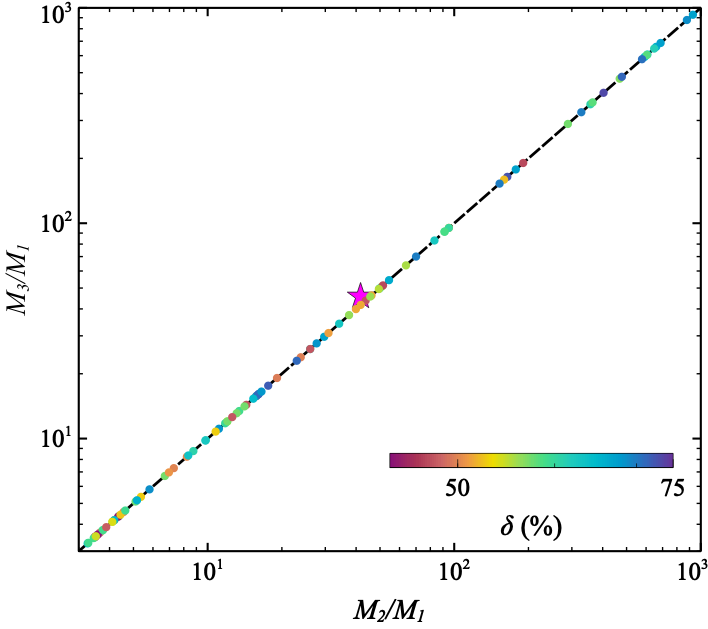}
\includegraphics[scale=0.42]{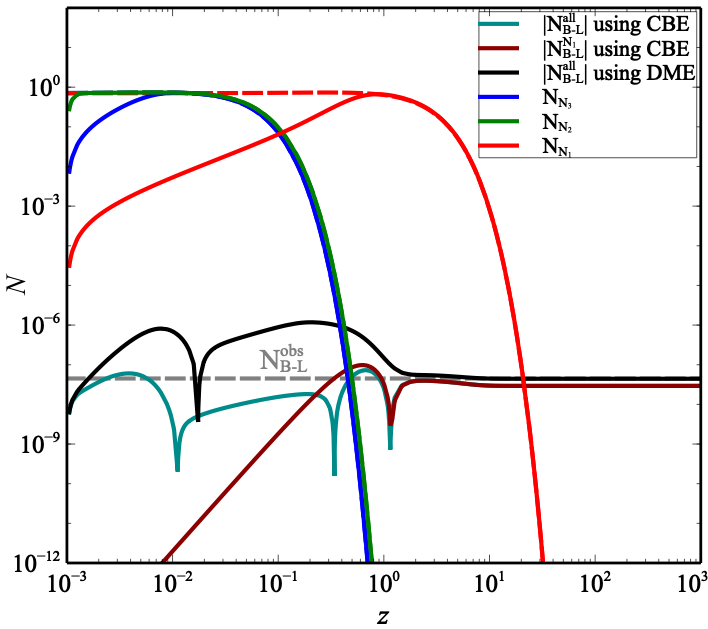}
    \caption{[\textit{Left}:] Allowed parameter space in the plane $M_3/M_1$ vs $M_2/M_1$ when $N_3$ is in weak washout regime. The color code depicts the $\delta$ values representing the \textit{memory} effect. [\textit{Right}:] Cosmological evolution of comoving number densities of RHNs and $B-L$ asymmetries for the magenta star point when $N_3$ is in the weak washout regime.}
    \label{fig:N3weakscattering}%
\end{figure*}

Now we turn to the final possibility, i.e., when $N_3$ is in the weak washout regime. The C.I. parameters are fixed as given in Table \ref{tab:cases} (case-4). Here, $K_1=7.95, \,K_2=46.44$ and $K_3=7.99\times 10^{-1}$. We fix the mass of $N_1$ to be $M_1=3\times10^{12}\rm{GeV}$. From Eq.(\ref{eq:N3weakProjection}), we get $\mathcal{P}^{(2)}_{11}=0.004$ and $\mathcal{P}^{(3)}_{11}=0.412$, which tell us that the decay of $N_2$ is completely orthogonal to $N_1$, whereas the decay of $N_3$ has both parallel and perpendicular components to $N_1$. We show points corresponding to the correct $B-L$ asymmetry and their memory effect in terms of mass ratios $M_2/M_1$ vs $M_3/M_1$ in the left panel of Fig. \ref{fig:N3weakscattering}. We observe that all points lie very close to the diagonal line, indicating that the masses of $N_2$ and $N_3$ are very close to each other. The region above these points corresponds to underabundance. In this scenario, $N_2$ is in a relatively strong washout regime, which suppresses both the asymmetry generated by $N_3$ and its own produced asymmetry. Consequently, the heavier generation can significantly impact the final $B-L$ asymmetry only when $M_3 \gtrsim M_2$, where it can generate a sufficiently large memory effect to yield the correct $B-L$ asymmetry.

This behavior is further explained in Fig.~\ref{fig:N3weakscattering}(\textit{right}), where we consider the benchmark mass ratios \( M_2/M_1 =42 \) and \( M_3/M_1 = 46 \), ensuring that the mass ratios are relatively close. The Yukawa coupling matrix for this benchmark point is given as
\begin{widetext}
    \begin{eqnarray}
        \label{eq:N3weakYukawa}
        Y=
        \begin{pmatrix}
(-1.42 \times 10^{-4} - 1.58 \times 10^{-2} i) & (-3.26 \times 10^{-2} + 5.54 \times 10^{-2} i) & (-1.54 \times 10^{-2} + 2.51 \times 10^{-2} i) \\
(-2.16 \times 10^{-3} - 1.89 \times 10^{-2} i) & (3.52 \times 10^{-3} - 3.10 \times 10^{-1} i) & (-1.11 \times 10^{-2} - 8.03 \times 10^{-3} i) \\
(-1.60 \times 10^{-3} + 1.51 \times 10^{-2} i) & (-3.24 \times 10^{-3} - 3.25 \times 10^{-1} i) & (1.35 \times 10^{-2} - 5.13 \times 10^{-2} i)
\end{pmatrix}.
    \end{eqnarray}
\end{widetext}

The color code remains the same as in previous cases. Since the masses of $N_2$ and $N_3$ are close, the asymmetry generated by $N_3$ is large and is not significantly affected by the washout effects of $N_2$. Finally, due to projection effects, this asymmetry is protected and adds to the asymmetry produced by $N_1$, resulting in the final $B-L$ asymmetry.

\section{Revisiting previous analytical results with full numerical solutions}\label{sec:x}
We now turn to compare the analytical solution of DME of earlier works \cite{Barbieri:1999ma,Engelhard:2006yg,BariDensityMatrix,N2leptogenesisprojection,Leptogenesisin2RHNModel,JessicaTurner,BariReview,DiBari:2005st,Hahn-Woernle:2009okg, ReFiorentin:2015rrc} to our numerical solution of the DME. In the limit of two RHNs, the analytical solution of Eq.(\ref{eq:Densitymatrixeq}) is given as \cite{BariDensityMatrix}
\begin{eqnarray}
    N^{B-L}_{\rm ana}=\varepsilon_1\kappa(K_1)+(e^{-\frac{3\pi}{8}K_1}p_{12}+1-p_{12})\varepsilon_2\kappa(K_2)\nonumber\\
\end{eqnarray}
where $\kappa(K_1)$ is the final efficiency factor given as
\begin{eqnarray}
    \kappa(x)=\frac{2}{xz_{\rm B}(x)}\left( 1-e^{-\frac{1}{2}xz_{\rm B}(x)} \right)
\end{eqnarray}
where,
\begin{eqnarray}z_{B_i}\approx2+4K_i^{0.13} e^{-2.5/K_i}.
\end{eqnarray}
\begin{figure}[h]
    \centering
    \includegraphics[scale=0.4]{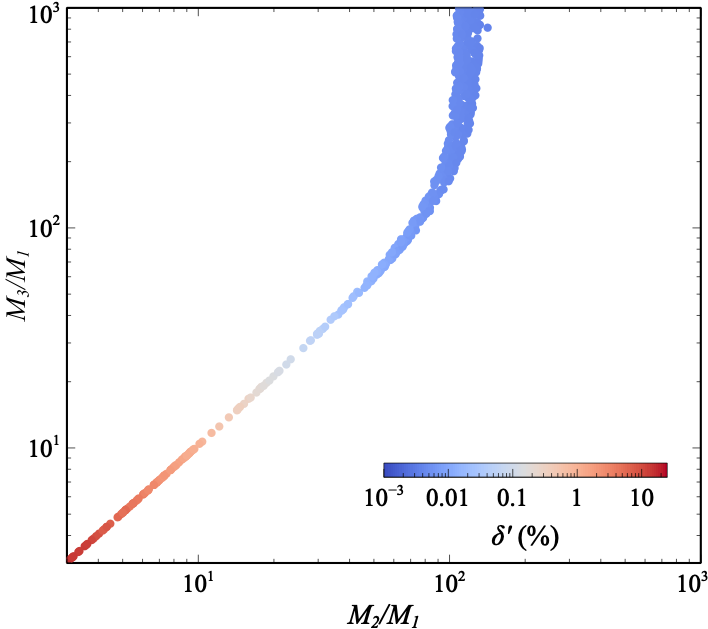}
    \caption{Correct lepton asymmetry parameter space in the plane of $M_3/M_1-M_2/M_1$ with $M_1=4\times10^{12}$ GeV and $\theta=5.317 \times 10^{-4}-1.685 \times 10^{-2}*i$. The color code denotes $\delta^\prime$. See main text for details.}
    \label{fig:masssplitting02}
\end{figure}
To quantify the validity of the analytical approximation to the density matrix equations (DMEs), we define the relative deviation parameter
\begin{equation}
\delta^\prime \equiv
\frac{N^{B-L}_{\rm num}-N^{B-L}_{\rm ann}}{N^{B-L}_{\rm ann}}\times100\% ,
\end{equation}
where $N^{B-L}_{\rm num}$, $N^{B-L}_{\rm ann}$
denote the lepton asymmetry obtained from the full numerical solution of the DMEs and from the analytical approximation, respectively. In Fig. \ref{fig:masssplitting02}, we present the parameter space yielding the correct baryon asymmetry in the plane of $M_3/M_1$ vs $M_2/M_1$ for $\theta=5.317\times10^{-4}-i1.685\times10^{-2}$ and $M_1=4\times10^{12}$ GeV. The color coding represents the magnitude of $\delta^\prime$. We observe that for large $M_2/M_1\simeq20-30$, analytical estimation matches very accurately with the numerical solution. On the other hand, for small $M_2/M_1\simeq 3-10$, the deviation typically lies in the range $\delta^\prime\gtrsim10-50\%$, indicating that the analytical approximation systematically underestimates the baryon asymmetry compared to the full numerical DME solution. This behavior is physically expected. When the mass hierarchy is small, $N_2$ remains thermally populated during the epoch in which $N_1$ dynamics are relevant. In that case, the off-diagonal entries of the density matrix contribute appreciably to the final asymmetry. For larger hierarchies, $N_2$ has already decayed or is Boltzmann suppressed well before $N_1$ becomes active, and the two-flavored analytical approximation matches the full numerical solution. This demonstrates the necessity of numerically accounting for projection effects in DME solutions, particularly in moderate mass hierarchies.

\section{Importance of projection effects in Probing leptogenesis with neutrinoless double beta decay}\label{sec:nu0bb}

Neutrinoless double beta ($0\nu\beta\beta$) decay constitutes one of the most sensitive probes of the Majorana nature of neutrinos and provides complementary information on the absolute neutrino mass scale. The quantity relevant for $0\nu\beta\beta$ decay experiments is the effective neutrino mass, defined as
\begin{eqnarray}
    m_{\beta\beta}=&& {m_1} \cos^2\theta_{12}\cos^2\theta_{13}e^{2i\phi_1} \nonumber\\&&+m_2 \sin^2\theta_{12}\cos^2\theta_{13}e^{2i\phi_2}+m_3 \sin^2\theta_{13}.\label{eq:mbetabeta}
\end{eqnarray}
Here, $m_1$ denotes the lightest SM neutrino mass. Assuming normal ordering, the remaining neutrino masses are given by $m_2=\sqrt{\Delta m_{21}^2+m_1^2}$ and $m_3=\sqrt{\Delta m_{31}^2+m_1^2}$, while $\phi_1$ and $\phi_2$ represent the Majorana phases. The lightest neutrino mass $m_1$ thus emerges as a common parameter governing both leptogenesis and neutrinoless double beta ($0\nu\beta\beta$) decay. In this work, we investigate the possibility of constraining leptogenesis through current and future sensitivities of $0\nu\beta\beta$ decay experiments.\footnote{For simplicity, and in order to clearly illustrate the impact of projection effects, we ignore the contribution of $N_3$ to the final lepton asymmetry.}
\begin{figure}[h]
    \centering
    \includegraphics[scale=0.4]{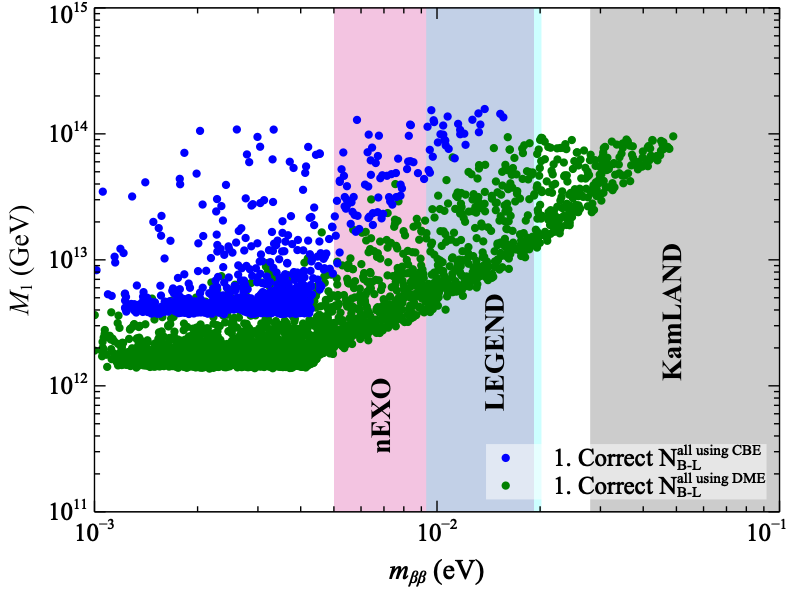}
    \caption{The lightest RHN mass as a function of effective neutrino mass. Here we fixed $\theta=3.38582 + 1.50555 \times 10^{-4} i$ radians, $M_2/M_1=25$ and $M_3/M_1=100$. The neutrino oscillation data are varied within $3\sigma$ range with the Majorana phases varied within $\left[0,2\pi\right]$. The KamLAND-Zen exclusion limit is shown as a gray-shaded region. The future sensitivities of nEXO and LEGEND are shown in pink and cyan shaded regions, respectively.}
    \label{fig:3sigma}
\end{figure}
Figure \ref{fig:3sigma}  shows the correlation between the lightest right-handed neutrino mass $M_1$ and the effective neutrino mass $m_{\beta\beta}$ relevant for neutrinoless double beta decay in three different scenarios. Here, the blue points correspond to the parameter choices that yield the observed baryon asymmetry when all the RHNs are considered for the asymmetry production with CBE, while the green points represent solutions obtained using the density matrix equations, which consistently account for projection effects. Here, we fix the CI rotation angle to $\theta=3.38582+i1.50555\times10^{-4}$, taking hierarchical heavy neutrinos with mass ratios $M_2/M_1=25$ and $M_3/M_1=100$, and vary the neutrino oscillation parameters in $3\sigma$ range with Majorana phases in the range $[0,2\pi]$ \cite{Esteban:2024eli}. The shaded regions indicate current and future sensitivities of neutrinoless double beta decay experiments: the present KamLAND-Zen \cite{KamLAND-Zen:2024eml} exclusion bound (gray), and the projected sensitivities of nEXO \cite{nEXO:2021ujk} (pink) and LEGEND \cite{LEGEND:2021bnm} (cyan). A clear separation is observed between the regions preferred by the Boltzmann and density matrix treatments. In particular, the density matrix approach allows successful leptogenesis for significantly lower values of $M_1$ at fixed $m_{\beta\beta}$, reflecting the impact of projection effects neglected in the classical approximation. Remarkably, a substantial portion of the parameter space compatible with density-matrix leptogenesis falls within the reach of upcoming neutrinoless double beta decay experiments, highlighting a non-trivial interplay between the dynamics of leptogenesis and low energy lepton number violating observables. The reason for this is that for this choice of parameters, $\ell_2$ and $\ell_1$ are not parallel to each other and hence have significant projection effects. These effects reduce the required mass scale of the lightest right-handed neutrino, thereby allowing the correct baryon asymmetry to be achieved for larger values of $m_1$ while keeping the Yukawa couplings within the perturbative regime. Such a parameter space is not accessible within the classical Boltzmann approach. This distinction is reflected in the predictions for neutrinoless double beta decay.

\section{Conclusions}\label{sec:conclu}

In the standard treatment of thermal leptogenesis with hierarchical right-handed neutrinos, it is commonly assumed that the final $B-L$ asymmetry is solely determined by the dynamics of the lightest state, $N_1$, since any pre-existing asymmetry generated by $N_2$ or $N_3$ is expected to be completely washed out. While several earlier studies have demonstrated that this assumption can fail in scenarios of $N_2$-dominated leptogenesis, such analyses typically rely on classical Boltzmann equations and specific washout hierarchies.
In this work, we have shown that a qualitatively different, more general phenomenon emerges when flavor projection effects are treated consistently within the density-matrix formalism. We find that the asymmetries generated by $N_2$ and $N_3$ are typically not fully aligned with the flavor direction washed out by $N_1$, leading to a continuous, quantifiable memory effect, even in parameter regions where $N_1$ interactions remain active. This effect is therefore distinct from the conventional notion of $N_2$-dominated leptogenesis.
By combining the density-matrix evolution with constraints from low-energy neutrino oscillation data, we further show that only one RHN can be in the weak-washout regime, which naturally organizes the parameter space into four distinct dynamical regimes. In all cases, we find that the memory effect enhances the surviving asymmetry compared to classical Boltzmann treatments that consider either only $N_1$ decay or sequential decays of all RHNs. Our results demonstrate that the standard reduction of hierarchical leptogenesis to an effective single-RHN description can fail even outside the usual $N_2$-dominated regime, highlighting the importance of density matrix effects in accurately determining the final baryon asymmetry. We also show that when the mass hierarchy between $N_1$ and $N_2$ is less than 15-20, the analytical approximation underestimates the baryon asymmetry compared to the full numerical solution of density matrix equations.

We find that incorporating projection effects into the density-matrix treatment significantly enlarges the viable parameter space, extending it into the sensitivity range of neutrinoless double beta decay experiments. As a result, these experiments can probe a substantial portion of the parameter space relevant for both the seesaw mechanism and leptogenesis, thereby providing meaningful and complementary constraints on the underlying neutrino mass–generation scale and the dynamics responsible for the baryon asymmetry of the Universe.

\noindent
\acknowledgments
P.K.P. would like to acknowledge the Ministry of Education, Government of India, for providing financial support for his research via the Prime Minister’s Research Fellowship (PMRF) scheme.
The work of N.S. is supported by the Department of Atomic Energy - Board of Research in Nuclear Sciences, Government of India (Ref. Number: 58/14/15/2021-BRNS/37220). We thank the anonymous Referee(s) of Physical Review D  for their constructive comments, which improved the manuscript.

\onecolumngrid

\appendix
\section{Projection Matrices for different cases:}
Using the Eq.(\ref{eq:changeofbasis}) and Eq.(\ref{eq:flavor_rotation}), we can write down the projection matrices for all three right-handed neutrinos in the $(\ell_1, \ell_1^{\perp^{(1)}},\ell_1^{\perp^{(2)}})$ basis. 

\subsection{Projection matrices for section \ref{sec:leptogenesis}}
Here, we present the projection matrices obtained for the parameters given in Table \ref{tab:benchmarks}. The following projection matrices are used when $\ell_1$ and $\ell_2$ are parallel to each other, demonstrated in Fig.\ref{fig:parallelbp}(left).
\begin{eqnarray}
    \label{eq:parallelProjection}
    \mathcal{P}^{(1)}&=&
    \begin{pmatrix}
        1&0&0\\
        0&0&0\\
        0&0&0
    \end{pmatrix};
    \mathcal{P}^{(2)}=
        \begin{pmatrix}
0.997 &
-0.011 - 0.038 i &
0.033 - 0.011 i \\
-0.011 + 0.038 i &
0.002 &
0.000 + 0.001 i \\
0.033 + 0.011 i &
0.000 - 0.001 i &
0.001
\end{pmatrix}.
\end{eqnarray}

The following projection matrices are used when $\ell_1$ and $\ell_2$ are orthogonal to each other, demonstrated in Fig.\ref{fig:parallelbp}(right).

\begin{eqnarray}
    \label{eq:orthogonlProjection}
    \mathcal{P}^{(1)}&=&
    \begin{pmatrix}
        1&0&0\\
        0&0&0\\
        0&0&0
    \end{pmatrix};
    \mathcal{P}^{(2)}=
\begin{pmatrix}
0.003 &
-0.008 + 0.045 i &
-0.025 - 0.0003 i \\
-0.008 - 0.045 i &
0.770 &
0.067 + 0.413 i \\
-0.025 + 0.0003 i &
0.067 - 0.413 i &
0.227
\end{pmatrix}
\end{eqnarray}

\subsection{Projection matrices for section \ref{sec:result}}
Here, we present the projection matrices obtained for each washout regime using the parameters given in Table \ref{tab:cases}.

\subsubsection{All Strong regime}
\textbf{Case-1 (a):}
\begin{eqnarray}
    \label{eq:allstrongProjection}
    \mathcal{P}^{(1)}&=&
    \begin{pmatrix}
        1&0&0\\
        0&0&0\\
        0&0&0
    \end{pmatrix};
    \mathcal{P}^{(2)}=
    \begin{pmatrix}
0.201 & 0.300 + 0.032 i & -0.026 + 0.262 i \\
0.300 - 0.032 i & 0.453 & 0.003 + 0.396 i \\
-0.026 - 0.262 i & 0.003 - 0.396 i & 0.346
\end{pmatrix};\nonumber\\
    \mathcal{P}^{(3)}&=&
    \begin{pmatrix}
0.633 & 0.253 + 0.260 i & -0.226 + 0.223 i \\
0.253 - 0.260 i & 0.208 & 0.001 + 0.182 i \\
-0.226 - 0.223 i & 0.001 - 0.182 i & 0.159
\end{pmatrix}.
\end{eqnarray}

\textbf{Case-1 (b):}
\begin{eqnarray}
    \label{eq:allstrongProjection2}
    \mathcal{P}^{(1)}&=&
    \begin{pmatrix}
        1&0&0\\
        0&0&0\\
        0&0&0
    \end{pmatrix};
    \mathcal{P}^{(2)}=
    \begin{pmatrix}
0.003 & -0.013 + 0.040 i & -0.026 - 0.005 i \\
-0.013 - 0.040 i & 0.709 & 0.046 + 0.450 i \\
-0.026 + 0.005 i & 0.046 - 0.450 i & 0.288
\end{pmatrix};\nonumber\\
    \mathcal{P}^{(3)}&=&
    \begin{pmatrix}
0.002 & -0.011 - 0.031 i & 0.019 - 0.009 i \\
-0.011 + 0.031 i & 0.710 & 0.046 + 0.450 i \\
0.019 + 0.009 i & 0.046 - 0.450 i & 0.288
\end{pmatrix}.
\end{eqnarray}

\subsubsection{$N_1$ in weak washout regime}

\begin{eqnarray}
    \label{eq:N1weakProjection}
    \mathcal{P}^{(1)}&=&
    \begin{pmatrix}
        1&0&0\\
        0&0&0\\
        0&0&0
    \end{pmatrix};
    \mathcal{P}^{(2)}=
    \begin{pmatrix}
0.059 & -0.194 - 0.081 i & 0.107 + 0.023 i \\
-0.194 + 0.081 i & 0.741 & -0.379 + 0.069 i \\
0.107 - 0.023 i & -0.379 - 0.069 i & 0.200
\end{pmatrix};\nonumber\\
    \mathcal{P}^{(3)}&=&
    \begin{pmatrix}
0.707 & -0.403 + 0.020 i & 0.204 - 0.048 i \\
-0.403 - 0.020 i & 0.230 & -0.118 + 0.022 i \\
0.204 + 0.048 i & -0.118 - 0.022 i & 0.062
\end{pmatrix}.
\end{eqnarray}

\subsubsection{$N_2$ in weak washout regime}

\begin{eqnarray}
    \label{eq:N2weakProjection}
    \mathcal{P}^{(1)}&=&
    \begin{pmatrix}
        1&0&0\\
        0&0&0\\
        0&0&0
    \end{pmatrix};
    \mathcal{P}^{(2)}=
    \begin{pmatrix}
0.008 & -0.065 + 0.002 i & -0.004 + 0.058 i \\
-0.065 - 0.002 i & 0.555 & 0.046 - 0.490 i \\
-0.004 - 0.058 i & 0.046 + 0.490 i & 0.438
\end{pmatrix};\nonumber\\
    \mathcal{P}^{(3)}&=&
    \begin{pmatrix}
0.310 & -0.331 + 0.101 i & 0.062 + 0.301 i \\
-0.331 - 0.101 i & 0.386 & 0.032 - 0.341 i \\
0.062 - 0.301 i & 0.032 + 0.341 i & 0.305
\end{pmatrix}.
\end{eqnarray}

\subsubsection{$N_3$ in weak washout regime}

\begin{eqnarray}
    \label{eq:N3weakProjection}
    \mathcal{P}^{(1)}&=&
    \begin{pmatrix}
        1&0&0\\
        0&0&0\\
        0&0&0
    \end{pmatrix};
    \mathcal{P}^{(2)}=
    \begin{pmatrix}
0.004 & -0.050 + 0.007 i & -0.006 - 0.032 i \\
-0.050 - 0.007 i & 0.707 & 0.023 + 0.452 i \\
-0.006 + 0.032 i & 0.023 - 0.452 i & 0.289
\end{pmatrix};\nonumber\\
    \mathcal{P}^{(3)}&=&
    \begin{pmatrix}
0.412 & 0.185 - 0.371 i & 0.243 + 0.106 i \\
0.185 + 0.371 i & 0.417 & 0.014 + 0.267 i \\
0.243 - 0.106 i & 0.014 - 0.267 i & 0.171
\end{pmatrix}.
\end{eqnarray}

\twocolumngrid
%

\end{document}